\documentclass[twocolumn]{aastex631}

\usepackage{graphicx}
\usepackage{rotating}
\usepackage{hyperref}
\usepackage{amsmath}
\newsavebox{\imagebox}

\newcommand{\nii}{[{\sc N\,ii}]}
\newcommand{\sii}{[{\sc S\,ii}]}

\begin{document}
\title{High Spectral Resolution Observations of the [SII] Emission Line Doublet in the Filamentary Nebula surrounding NGC 1275}

\author[0000-0002-2478-5119]{Benjamin Vigneron}
\affiliation{Département de Physique, Université de Montréal, Succ. Centre-Ville, Montréal, Québec, H3C 3J7, Canada}

\author[0000-0001-7271-7340]{Julie Hlavacek-Larrondo}
\affiliation{Département de Physique, Université de Montréal, Succ. Centre-Ville, Montréal, Québec, H3C 3J7, Canada}
\affiliation{Centre de recherche en astrophysique du Québec (CRAQ)}

\author[0000-0003-2001-1076]{Carter Lee Rhea}
\affiliation{Département de Physique, Université de Montréal, Succ. Centre-Ville, Montréal, Québec, H3C 3J7, Canada}
\affiliation{Centre de recherche en astrophysique du Québec (CRAQ)}
\affiliation{Dragonfly Focused Research Organization, 150 Washington Avenue, Santa Fe, 87501, NM, USA}


\author[0000-0002-7275-3998]{Christoph Pfrommer}
\affiliation{Leibniz-Institut
für Astrophysik Potsdam, An der Sternwarte 16
14482 Potsdam}





\author[0000-0003-2630-9228]{Greg L. Bryan}
\affiliation{Department of Astronomy, Columbia University, 550 West 120th Street, New York, NY 10027, USA}
\affiliation{Center for Computational Astrophysics, Flatiron Institute, 162 5th Avenue, New York, NY 10010, USA}

\author[0000-0002-2808-0853]{Megan Donahue}
\affiliation{Department of Physics and Astronomy, Michigan State University, East Lansing, MI 48824, USA}

\author[0000-0002-3398-6916]{Alastair Edge}
\affiliation{Department of Physics, Durham University, South Road, Durham DH1 3LE, UK}

\author[0000-0002-9378-4072]{Andrew Fabian}
\affiliation{Institute of Astronomy, University of Cambridge, Madingley Road, Cambridge CB3 0HA, UK}




\author[0000-0001-5226-8349]{Michael McDonald}
\affiliation{Kavli Institute for Astrophysics and Space Research, MIT, Cambridge, MA 02139, USA}


\author[0000-0001-7597-270X]{Annabelle Richard-Lafferrière}
\affiliation{Institute of Astronomy, University of Cambridge, Madingley Road, Cambridge CB3 0HA, UK}

\author[0000-0002-5136-6673]{Laurie Rousseau-Nepton}
\affiliation{Department of Astronomy and Astrophysics, University of Toronto, 50 St Georges Street West, Alexandria, ON K0C 1A0, Canada}

\author[0000-0002-3514-0383]{G. Mark Voit}
\affiliation{Department of Physics and Astronomy, Michigan State University, East Lansing, MI 48824, USA}


\author[0000-0003-0392-0120]{Norbert Werner}
\affiliation{Department of Theoretical Physics and Astrophysics, Faculty of Science, Masaryk University, Brno, Czech Republic}



\begin{abstract}
We analyze new high-spectral resolution SITELLE observations (R = $\lambda/\Delta\lambda$ = 7000) of the filamentary nebula surrounding NGC 1275, central galaxy of the Perseus cluster.  We present here analysis of the \sii$\lambda6716$ and \sii$\lambda6731$ emission line doublet, using its ratio to determine the electron density of the optically emitting filaments. We compare these measurements with electron densities derived from deep Chandra X-ray observations of the intra-cluster medium (ICM) to determine if any correlations in density can be found. We report the detection of a clear dichotomy between the outer filaments, displaying on average lower \sii\text{ }emission line ratio of $\sim 1.1$ and the inner filaments displaying higher ratios of $\sim 1.3$. These results indicate that most of the gaseous filaments lie close to the low-density threshold for the density measurement of $\sim 10^2\text{ cm}^{-3}$.  Using radial profiles, we find that the inner filaments have a roughly constant density, whereas the ICM density decreases with radius. In the outer filaments, we observe hints of local connections between the densities of the ICM and optical filaments, but no clear correlation seems to be observed overall. We also combined these density measurements with cold molecular CO gas observations to derive a relationship between temperature, density and pressure for the multiphase environment surrounding NGC 1275. Finally, we investigated potential models to explain the observed density measurements and explored similar studies of filamentary nebula around other central galaxies of cool-core galaxy clusters. 

\end{abstract}




\section{Introduction} \label{sec:intro}

Galaxy clusters are the largest gravitationally virialized structures in the Universe, containing hundreds to thousands of galaxies (e.g. \citealt{bahcall_clusters_1977}, \citealt{abell_catalog_1989}). Galaxies within the cluster only represent a very small fraction of its mass ($\sim 5\%$), while dark matter is predominant ($\sim 80\%$ - e.g. \citealt{sand_dark_2004}, \citealt{voigt_galaxy_2006}). Finally, the intra-cluster medium or ICM is the other main component of the cluster in terms of mass ($\sim 15\%$) and is composed of plasma at extremely high temperature ($\sim 10^7-10^8$ K), but low electron density overall ($\sim 10^{-1}-10^{-3} \text{ cm}^{-3}$ in the core and with a mean cluster gas density of $1.5\times10^{-4}\text{ cm}^{-3}$- e.g. \citealt{cavaliere_extragalactic_1971}, \citealt{gursky_detection_1971}). Due to its extremely high temperature and since it is predominantly composed of ions and free electrons, the ICM emits brightly in X-rays mainly through the bremsstrahlung effect with powerful Fe forbidden lines as well as many lower-energy lines from ions, including those of C, N, O, Ne, Mg, Si, S, Ar, and Ca (\citealt{sarazin1986x}).\\

The
surface brightness of the ICM as a function of distance from the cluster's center has often been the basis of their classification: cool-core clusters have a strongly peaked X-ray emission profile, while non-cool core clusters display a more homogeneous and diffuse X-ray profile (\citealt{million2009chandra}, \citealt{hudson2010cool}). Many galaxy clusters have a dominant central galaxy called the brightest cluster galaxy (BCG), which frequently exhibits an extended filamentary optical nebula of ionised gas in the case of cool-core clusters (\citealt{crawford_rosat_1999}, \citealt{mcdonald_origin_2010}). The filaments are highly multiphase, generally displaying high H$\alpha$ luminosities up to a few $10^{42}\text{ erg~s}^{-1}$ (e.g. \citealt{conselice2001nature})\\

In this paper, we analyze the filamentary nebula of ionised gas surrounding NGC 1275 ($z = 0.017284$, Hitomi \citealt{hitomi_collaboration_atmospheric_2018}), the BCG of the nearby Perseus cluster of galaxies, which has been studied extensively at various wavelengths (see e.g. \citealt{forman_observations_1972}, \citealt{fabian_wide_2011}, \citealt{salome_very_2011}, \citealt{lim_molecular_2012}, \citealt{nagai_alma_2019}). Moreover, this cluster displays several X-ray cavities generated by a succession of radio jets emitted by the supermassive black hole (SMBH) at the center of the active galactic nucleus (AGN) of NGC 1275 (\citealt{1993MNRAS.264L..25B}, \citealt{fabian_chandra_2000}, \citealt{fabian_wide_2011}, \citealt{gendron-marsolais_revealing_2018}). These jets have displaced the ICM and fueled buoyantly rising radio-emitting bubbles (see \citealt{dunn2005radio}), which in turn have injected a large quantity of energy within the surrounding ICM through shocks, turbulence and mixing (e.g. \citealt{graham_weak_2008}, \citealt{randall_very_2015}, \citealt{zhuravleva_turbulent_2014}), mostly preventing its cooling as well as intense star formation in the BCG (\citealt{crawford_rosat_1999}, \citealt{best_prevalence_2007}, \citealt{voit_conduction_2008}). Nevertheless, the extended filamentary structures surrounding BCGs are believed to be the result of thermal instabilities generated by feedback from the AGN, leading to local cooling of the ICM which precipitates into denser clouds, thus funneling more material to the AGN. This self-regulated loop is described through the so-called precipitation model and other similar feedback models such as chaotic cold accretion or stimulated feedback (\citealt{1989agna.book.....O}, \citealt{gaspari2013chaotic}, \citealt{voit2015cooling}, \citealt{mcnamara_mechanism_2016}). Recent analyses indicate that some cooling might still be occurring in the form of hidden cooling flows and results in the formation of cold clouds as well as limited star formation (\citealt{fabian2022hidden}).\\

The filamentary nebula surrounding NGC 1275 is one of the largest ever observed with a size of $80$ kpc $ \times \text{ } 50$ kpc (e.g. \citealt{mcnamara_optical_1996}, \citealt{conselice2001nature}, \citealt{hamer_optical_2016}). The first observations of these gaseous filaments made by \cite{minkowski_optical_1957}, \cite{lynds1970improved}, and \cite{rubin1977new} separated the emission structure into a high-velocity (HV) feature ($\sim$ 8200 km/s) corresponding to a foreground spiral galaxy infalling onto NGC 1275 (\citealt{2015ApJ...814..101Y}, \citealt{rhea2025mapping}), and a low-velocity (LV) structure ($\sim$ 5200 km/s) associated with the emission of ionised gas encircling NGC 1275. Subsequent \textit{Hubble Space Telescope} observations of the LV structure then revealed the filamentary appearance of the ionised gas (\citealt{fabian_magnetic_2008}), while soft X-ray counterparts were discovered for certain bright filaments with the \textit{Chandra X-Ray Observatory} (\citealt{fabian_relationship_2003}).\\

Moreover, previous Chandra observations of the ICM surrounding NGC 1275 allowed the electron density of the X-ray emitting plasma to be determined.  \cite{fabian2011energy} found the average electron density of the hot ICM surrounding the filaments to be $\sim 0.035 \text{ cm}^{-3}$. They also note a clear drop in ICM temperature and increase in its density coincident with the optical filaments. Similarly, \cite{fabian2000chandra} obtained ICM pressure maps derived from the product of temperature and electron density around NGC 1275. These maps show a trend of an increased electron density around the central galaxy and subsequently diminishing radially, which is a clear feature of cool-core galaxy clusters.\\

The proximity of the Perseus cluster makes NGC 1275 a prime target to study the various properties of filamentary nebula around BCGs such as their formation and ionization mechanisms, as well as their electron densities. These can be determined by studying the effects of collisional de-excitation of specific forbidden emission lines such as the optical \sii$\lambda6716$ and \sii$\lambda6731$ emission line doublet. The ratio of this doublet is sensitive to the electron density, but not the temperature, because the excitation energies are close to one another (\citealt{wang2004reexamination}, \citealt{1989agna.book.....O}). \cite{heckman1989dynamical} attempted the first studies regarding the \sii$\lambda6716$ and \sii$\lambda6731$ emission line doublet ratio across the filamentary nebula surrounding NGC 1275, using a single measurement of \sii\text{} emission line ratio through slit spectroscopy, which only covered  the central region of the filaments ($r<1$ kpc). Nevertheless, they offered the first glimpse of the optical electron density of the filaments and their comparison with X-ray densities. \cite{heckman1989dynamical} also derived the gas pressure based on the optical electron density obtained from the \sii\text{} emission line ratio and demonstrated a higher pressure close to the central galaxy, which then decreases by factors of 2-4 at radii of 2-5 kpc.\\

Following this, a dedicated study of optical emission line ratios from the filamentary nebula surrounding NGC 1275 was produced by \cite{sabra2000emission}, using slit spectroscopy over four slits covering the central eastern and western regions of the filaments over a distance of $\sim 10 \text{ to } 20$ kpc. This study included an analysis of \sii\text{} emission line ratios as well as their associated electron density across the slits. The authors determined that line ratios would lead to an electron density below $n_e = 10^2\text{ cm}^{-3}$.\\

Similar analyses were performed in more detail by \cite{hatch_origin_2006}, where slit spectroscopy was used with six slits positioned along filaments. This allowed the measurement of \sii\text{} emission line ratios across inner filaments as well as extended outer north-western filaments up to $\sim 50$ kpc away from the galaxy and produced the first ratio profile as a function of H$\alpha$ surface brightness. This profile demonstrated that the optical electron density across the filaments remained at or below the low-density threshold of $n_e = 10^2\text{ cm}^{-3}$. \\

However, these studies leave a caveat as the \sii\text{} emission line ratio across the entirety of the filamentary nebula has not been determined. Doing so would allow us to determine potential spatial correlation of electron density with notable features such as cavities, X-ray filaments or shock fronts, which can be observed in X-ray through the ICM emission (see Fig. \ref{fig:fov}). Furthermore, recent studies revealed a strong correlation in surface brightness between optical and X-ray emitting filaments around BCGs hinting at a possible similar excitation mechanism (see \citealt{olivares2025halpha}). The SITELLE (\textit{Spectromètre Imageur à Transformée de Fourier pour l’Etude en Long et en Large de raies d’Emission}) Fourier transform imaging spectrometer at the Canada France Hawaii Telescope (CFHT) is an instrument of choice for such study, thanks to its extended field of view of 11' by 11'. This allows us to observe and retrieve the emission spectra at optical wavelengths of the entire filamentary nebula surrounding NGC 1275. A detailed study by \cite{gendron-marsolais_revealing_2018} of past SITELLE observations at low spectral resolution of R = 1800 was unable to properly separate and analyze the \sii\text{} emission lines across the filaments, which motivated renewed observations at higher spectral resolution of R = 7000 (see also \citealt{2024ApJ...962...96V}).\\

In this paper, we study these new high-spectral resolution observations (R = 7000) of the filamentary nebula surrounding NGC 1275 obtained with SITELLE. Thanks to the high-spectral resolution and extended field-of-view of SITELLE, we are able to map the \sii\text{} ratio in the outer filaments as well as in the central region in great detail therefore allowing us to perform a dedicated study of the electron density across the entirety of the filamentary nebula. We also compare these results to electron densities derived from deep Chandra X-ray observations of the permeating ICM, with a total exposure time of 808.4 ks.\\

In Section 2, we present SITELLE optical observations and the analysis of the \sii\text{ }emission line doublet, followed by a study of  Chandra X-ray observations and the electron densities derived from them. In Section 3 we present our results regarding the optical \sii\text{} emission line doublet ratio and derived electron density in the optical. Finally, a discussion of our results will be carried out in Section 4.\\

To directly compare our results to those of \cite{gendron-marsolais_revealing_2018} and the Hitomi \cite{hitomi_collaboration_atmospheric_2018}, we also adopt for NGC 1275 a redshift of $z=0.017284$, which implies an angular scale of $21.2 \; \text{kpc arcmin}^{-1}$. This redshift also corresponds to a luminosity distance of $75.5$ Mpc, assuming $H_0 = 69.6 \; \text{km s}^{-1} \text{Mpc}^{-1}$, $\Omega_M = 0.286$ and $\Omega_{vac} = 0.714$.


 %

\section{Data Reduction and Analysis} \label{sec:data}

\subsection{Observations with SITELLE}

\begin{figure*}
    \centering\includegraphics[width=175mm,scale=0.5]{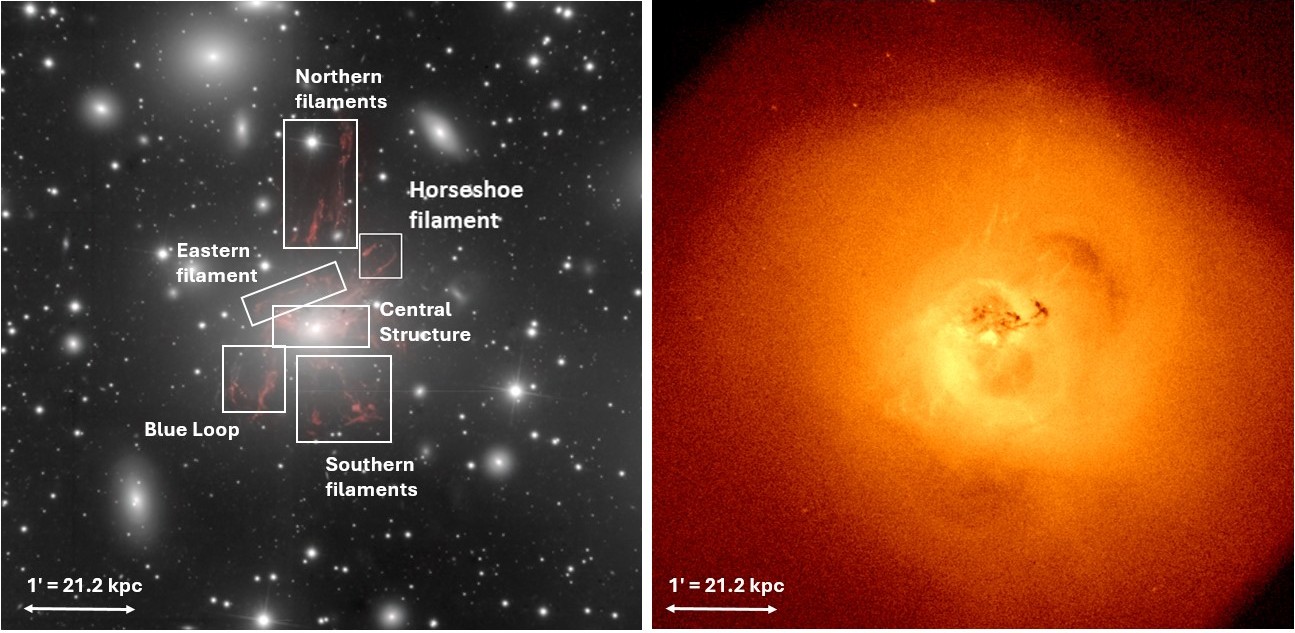}
    \caption{Left: SITELLE observations of NGC 1275 using the SITELLE SN3 filter (648-685 nm) at high-spectral resolution (R = 7000) and with a field of view of 11 arcmin by 11 arcmin. We display here the integrated flux images in greyscale with the brightest H$\alpha$ filaments in red. The white square and rectangles indicate specific regions and filaments of relevance. Right: Deep Chandra X-ray observations of NGC 1275 between 500 eV to 1 keV, displaying the soft X-ray counterparts of the optical emission line filaments (private communication).}
    \label{fig:fov}
\end{figure*}

SITELLE is a Fourier transform imaging spectrometer possessing an incredibly large field of view of 11' by 11' and installed with two E2V detectors of 2048 × 2064 pixels. This combination results in a spatial resolution of 0.321 × 0.321 arcsecs.

SITELLE was used in February 2020 to observe the filamentary nebula surrounding NGC 1275 during Queued Service Observations 20AD99 by PIs Hlavacek-Larrondo and Rhea (see Figure \ref{fig:fov}). To obtain the \sii\text{} emission line doublet studied here, the SN3 filter of SITELLE was used to cover a wavelength range from 648 nm to 685 nm. These observations were centered on NGC 1275 with RA 03:19:48.16 and DEC +41:30:42.1 and achieved an extremely high spectral resolution of $R=7000$ (corresponding to a resolution of 42.8 km/s or 0.938 \AA) after 4 hours (1710 exposures of 8.42s) of observations including overheads.

Five emission lines are covered with the SN3 filter, namely: \sii$\lambda6716$, \sii$\lambda6731$, \nii$\lambda6584$, H$\alpha$(6563\AA), and \nii$\lambda6548$. The kinematic analysis of these emission lines was presented in \cite{2024ApJ...962...96V}. In this paper, we mostly focus on the doublet of ionised sulfur emission lines \sii$\lambda6716$ and \sii$\lambda6731$.

The data reduction of SITELLE is performed at CFHT through a dedicated pipeline: ORBS, which we will describe briefly here. First, a correction of the electronic bias and flat-field curvature of the interferometric images is realised. Following this step, images are aligned to solve potential guiding errors that could have occured. Afterwards, a detection of potential cosmic rays hitting the instrument optics is performed by comparing successive observations to determine if any abnormal flux increase can be detected for a given pixel. To correct these anomalies linked to cosmic rays, an estimation of the gaussian flux of neighboring pixels is determined and applied on the affected pixel. Moreover, atmospheric variations during observations are taken into consideration and corrected by using a transmission function obtained while acquiring data. After these steps have been performed, the reduction pipeline will produce a Fourier transform of all the interferograms, which are then phase corrected. Finally, wavelengths and flux calibrations are realized to adequately record the high-spectral resolution observations (\citealt{2019MNRAS.485.3930D}).

\subsubsection{[SII] Emission Line Fitting}
\label{sii_fitting}

A common difficulty with the study of the \sii\text{} emission line doublet in filamentary nebulae surrounding BCGs is its relatively low flux compared to other prominent optical emission lines such as H$\alpha$ or \nii\text{}. Therefore, to improve the detection of the \sii\text{} emission line doublet, we used several mitigation techniques outlined below in order to obtain the most precise flux measurement for these emission lines.

An additional difficulty with the study of the \sii\text{} emission lines within the SITELLE data lies in the fact that the background spectral emission severely overlaps with the emission line doublet. This background emission is mainly produced by galaxies near the optically emitting filaments as well as the diffuse emission of other structures in the field-of-view. Therefore, in order to retrieve the \sii\text{ }emission lines, a background subtraction is necessary. However, since the SITELLE observations possess a high spectral resolution (R = 7000), the background emission shows a certain level of variability across the field of view (see \citealt{2024ApJ...962...96V} for more details). In order to tackle this issue, we decided to spatially bin multiple pixels into significantly larger regions within the outer and fainter filaments from which one emission spectrum would be extracted and fitted. This procedure also increases the overall flux from the \sii\text{ }emission line doublet within the outer filaments, reducing the impact of background subtraction on the doublet. We defined these larger regions by taking into account the morphology of the filaments, their kinematics, as well as the overall flux of the \sii\text{} emission line doublet within each region. Overall, we binned the outer filaments into 18 regions, which are illustrated as red rectangles in the top left corner of Fig. \ref{fig:maps_outer_filaments}. Examples of background-subtracted and fitted spectra can also be found in Appendix \ref{sii_fitted_spectra}.

In order to obtain the flux of the \sii \text{ }doublet, we used the new fitting analysis software called \texttt{LUCI} developped recently for the SITELLE instrument (\citealt{ rhea_novel_2020, rhea_machine-learning_2020, rhea_luci_2021, rhea_machine-learning_2021}). \texttt{LUCI}'s goals and capabilities are explained in great detail at \href{https://crhea93.github.io/LUCI/index.html}{https://crhea93.github.io/LUCI/index.html} (\citealt{rhea_luci_2021}). Using the \texttt{pyregion} package, we created masks based on the large regions covering the filaments we described previously. Afterwards, we used the \texttt{fit\_spectrum\_region} function of \texttt{LUCI} to specifically fit the spectra of the SITELLE pixels belonging to these regions while also performing a background removal based on the spectra extracted from smaller regions found in the field in close proximity to the regions fitted. We thus obtained the flux measurements of the \sii\text{ }emission line doublet for all regions which were then used to determine their ratios. Finally, we also determined the errors on our fits by using the Bayesian analysis implemented within \texttt{LUCI} which produces a complete Bayesian \textit{Monte-Carlo Markov Chain} (MCMC) approach using the python module \texttt{emcee}.

Since the central region displays significantly higher fluxes (up to an order of magnitude more than the outer filaments, see \citealt{2024ApJ...962...96V}), we decided to isolate it by defining the outer filaments to have fluxes below $\sim 1\times10^{-18} \text{ erg s}^{-1}\text{cm}^{-2}\text{\AA}^{-1}$. We were then able to conduct a radial analysis, as well as a dedicated flux ratio map of the \sii\text{} emission line doublet for this central structure, allowing us to better understand the radial dependence of the ratio and by extension the density of the gas. Indeed, the emission properties of the central region differ significantly from those of the outer filaments (\citealt{2024ApJ...962...96V}). To perform this analysis, we used the weighted Voronoi tessellation (WVT) algorithm of \texttt{LUCI} which creates bins of pixels with a SNR threshold of 30 (see \citealt{2024ApJ...962...96V} and \href{https://crhea93.github.io/LUCI/index.html}{https://crhea93.github.io/LUCI/index.html} for a description of this methodology). The resulting bins are shown in the top left panel of Fig. \ref{fig:maps_central_structure}. Following this, we can use the \texttt{wvt\_fit\_region} function of \texttt{LUCI} to fit the emission spectrum of each bin. After obtaining the resulting \sii$\lambda6716$ and \sii$\lambda6731$ flux maps, the flux ratio is obtained by dividing the two, giving the result shown in the top left panel of Fig \ref{fig:maps_central_structure}.

Finally, we also carried out a radial analysis of the \sii\text{} emission line ratio across the central structure by defining 25 annuli of $1.16$'' widths and centered on the AGN at coordinates of RA = 3:19:48.189 and DEC = +41:30:42.135, as shown in the top right panel of Fig \ref{fig:maps_central_structure}. To determine the \sii\text{} ratios for this radial analysis, we considered the average of the fitted ratio determined from the WVT algorithm within each annulus. Since the inner region displays a significant radial velocity variation (see \citealt{2024ApJ...962...96V}), extracting a single spectrum for each annulus resulted in blended and noisy emission lines which proved difficult to effectively fit. We decided instead to fit the emission spectra extracted from the smaller binned regions created through the WVT algorithm to derive their respective ratios, and determined afterwards the average ratio within each annulus, which are presented in the bottom left panel of Figure \ref{fig:maps_central_structure}. 

We stress, however, that the three dimensional structure of the filamentary nebula is intrinsically unknown, despite the high resolution spectroscopy, meaning that measurements of electron density should be interpreted with nuance and caution.



%

\subsection{Observations with the Chandra X-Ray Observatory}

In order to directly compare the optical electron density of the filamentary nebula with the density of its surrounding ICM, we used archival Chandra X-ray spectroscopic observations of the Perseus Cluster and fitted the spectra to extract gas densities.

To obtain adequate signal-to-noise ratio, we combined the 8 Chandra ACIS-S exposures centered on the Perseus cluster with the highest exposure times. The ObsIDs of all the observations used are listed in Table \ref{tab:1} as well as their corresponding exposure times and coordinates. Moreover, these ObsIDs were taken prior to the Chandra X-ray Telescope's low energy response reduction due to an accumulation of molecular layers. This contaminant severely attenuates low-energy X-rays and started to accumulate rapidly between 2015 to 2017 (\citealt{o2017modeling, plucinsky2018complicated}).

After gathering all of these observations, we first ran the robust data cleaning pipeline implemented within the \texttt{Pumpkin} software (\citealt{rhea_novel_2020}). To do this, we ran the \texttt{FaintCleaning} function allowing us to create background subtracted images for every observation we gathered, in order to obtain a merged image. The backgrounds for each ObsId are determined by the user from the CCD that is most distant from the cluster center as displayed during the data cleaning pipeline. Afterwards, cleaned evt2 files are generated for each observation and merged images are produced by combining all the gathered ObsIds resulting in a total exposure time of 808.4 ks.

\subsubsection{X-ray Emission Line Fitting}

We then proceeded with the fitting of the X-ray spectroscopic observations using the X-ray fitting pipeline \texttt{AstronomyTools} implemented within \texttt{Pumpkin}, a general purpose software tool for fitting X-ray spectra developped to make thermodynamic maps and line plots (\citealt{rhea_novel_2020}). Since we considered several Chandra X-ray observations within our analysis, we merged them to extract the resulting spectra and fitted them afterwards.

Our goal is to understand how the electron density of the hot X-ray emitting ICM correlates with the same property of the cool optically emitting filaments. We thus used the same outer filaments regions utilized for the \sii\text{} emission line ratio analysis, as described in section 2.1.1 and displayed in Figure \ref{fig:maps_outer_filaments}, to determine the density of the ICM in these regions.

First, we defined our model to fit the Chandra X-ray spectra. We used an absorbed thermal model (\texttt{PHABS*APEC}) for the ICM emission and determined the foreground column density of the Perseus Cluster as $N_\mathrm{H} = 0.135\times10^{22}~\mathrm{cm}^{-2}$ using the HEASARC column density calculation tool. Moreover, \texttt{Pumpkin} uses a standard background model containing both local X-ray emission, background AGN, and diffuse background emission which we use to fit the data using a C statistic. We thus model the background X-ray emission with two specific emission components: a soft X-ray galactic \texttt{APEC} model with a temperature of 0.18 keV and metallicity $Z=1$ and a hard cosmic X-ray \texttt{BREMSS} model with a temperature of $kT$ = 40 keV. Finally, metallicity is set as a free parameter during the fitting procedure.

We then used the \texttt{Fitting} function of the \texttt{AstronomyTools} package. This function allows us to convolve the emission model with the response matrix of the Chandra X-ray observatory and simultaneously fit the emission spectra of all the pixels found in the annuli we defined. From this, several emission parameters such as temperature, pressure, and density can be extracted using a deprojection step which we will describe afterwards.

Similarly, to study the potential correlation with the central structure seen in the optical with SITELLE, we considered the same radial annuli, as seen in Fig. \ref{fig:maps_central_structure}, used for its analysis with \texttt{LUCI}. 
In this case however, we use the \texttt{Fitting\_Deprojected} function of the \texttt{AstronomyTools} package to properly deproject the several annuli onto the Chandra X-ray observations. On the other hand, for the outer regions, we cannot use this deprojection methodology, since it requires spherical symmetry, which does not hold for the regions shown in Fig. \ref{fig:maps_outer_filaments}. Therefore, to work around this issue, we decided to use the norm of our X-ray spectral fitting divided by the pixel area of the regions used as a measure of the integral of the electron density squared along our line of sight, as will be explained in Section \ref{X-ray_electron_density}.






\begin{table*}[t]
\centering
\begin{tabular*}{\textwidth}{@{\extracolsep{\fill}}c*{5}{>{$}c<{$}}}
    \hline
ObsID & \text{Observation Date} & \text{Exposure Time (ks)} & \text{Pointing RA} & \text{Pointing DEC} \\
    \hline
        3209&                   08-08-2002&                  95.8&             03 19 47.60&              +41 30 37.00\\
        4289&                     10-08-2002&                  95.4&             03 19 47.60&              +41 30 37.00\\
        4948&                    09-10-2004&                  118.6&             03 19 48.20&              +41 30 42.40\\
        4950&                    12-10-2004&                  96.9&             03 19 48.20&              +41 30 42.40\\
        4951&                   17-10-2004&                  96.1&             03 19 48.20&              +41 30 42.40\\
        4952&                    14-10-2004&                  164.2&             03 19 48.20&              +41 30 42.40\\
        6139&                    04-10-2004&                  56.4&             03 19 48.20&              +41 30 42.20\\
        6145&                   19-10-2004&                  85.0&              03 19 48.20&              +41 30 42.20\\
        \hline
\end{tabular*}
\caption{\textit{Chandra} X-ray Observatory observations of the Perseus Cluster of galaxies used in this analysis. The first column indicates the ObsIDs of these observations followed by their dates. We chose these specific observations since they were the 8 having the highest exposure times as specified in the third column. Finally, we indicate the RA and DEC of the observations in the fourth and fifth columns.}
\label{tab:1}
\end{table*}



\section{Results} \label{sec:results}

\subsection{[SII] Emission Line Ratio}
\subsubsection{Outer filaments}

The resulting line ratio map for the outer filaments is displayed at the top right corner of Figure \ref{fig:maps_outer_filaments}, while the corresponding values for \sii\text{} ratios and electron densities of each regions are presented in Table \ref{tab:2}. After obtaining the flux of the \sii\text{ }emission lines doublet across the regions considered, we effectively divided their values to produce their ratio. As can be seen at the top right corner of Figure \ref{fig:maps_outer_filaments} and in Table \ref{tab:2}, most of the regions have line ratios close to or above $\sim 1.0$. Interestingly, a portion of the region encompassing the eastern filament, as seen in Fig. \ref{fig:fov}, shows a significantly lower ratio at around $\sim 0.8$.

\subsubsection{Central Structure}

We also explored the radial \sii\text{ }emission line ratio of the central bright and turbulent region as detected in \cite{2024ApJ...962...96V} and displayed in Fig. \ref{fig:fov}. The emission line ratio map of the central region, the annuli used as well as the \sii\text{ }emission line ratio and corresponding electron density values are all displayed in Figure \ref{fig:maps_central_structure}. This ratio map and the corresponding radial analysis showcase an overall slightly higher \sii\text{ }ratio at around $\sim 1.34$ across the central structure compared to the outer filaments. 

\subsubsection{Dichotomy between Inner and Outer Filaments}

Interestingly, \sii\text{ }emission line doublet ratio analyses carried out for gaseous filaments around other BCGs show similar trends where the ratios appear generally high (see for instance \citealt{farage2010optical}, \citealt{mcdonald2012optical}, \citealt{hamer_optical_2016}, \citealt{iani2019inquiring}, \citealt{ciocan2021vlt}, \citealt{tamhane2023radio}). In order to emulate the results and plots presented in \cite{mcdonald2012optical} and \cite{hamer_optical_2016} for instance, we created Figures \ref{fig:ratio_distribution} and \ref{fig:flux_distribution}, which we will describe below. 

We display in the left panels of Figure \ref{fig:ratio_distribution} the distribution of the \sii\text{} ratios for the central region, outer filaments and nucleus centered on NGC 1275. These distributions are color-coded and their corresponding regions are displayed in the right side of Figure \ref{fig:ratio_distribution}. First, in the top left panel, we plot the \sii\text{} ratio distribution for the central structure. In the second row, we plot the same distribution but for the outer filaments and finally, in the bottom left panel, we isolate the emission within the nucleus defined as a region with radius of $3.23$'' and centered on the AGN located at RA = 3:19:48.189 and DEC +41:30:42.135. As can be seen, line ratios for the central structure are centered around $\sim 1.34$ with a standard deviation of $0.098$, while values for the outer filaments are clustered around $\sim 1.1$ with a standard deviation of $0.17$. A dichotomy between the outer and inner filaments can then clearly be observed when comparing the top and middle left panels of Figure \ref{fig:ratio_distribution}. We further discuss these results and their interpretation in terms of electron density in Section \ref{section:discussion}. 

\begin{figure*}
\centering
\savebox{\imagebox}{\includegraphics[width=0.45\textwidth]{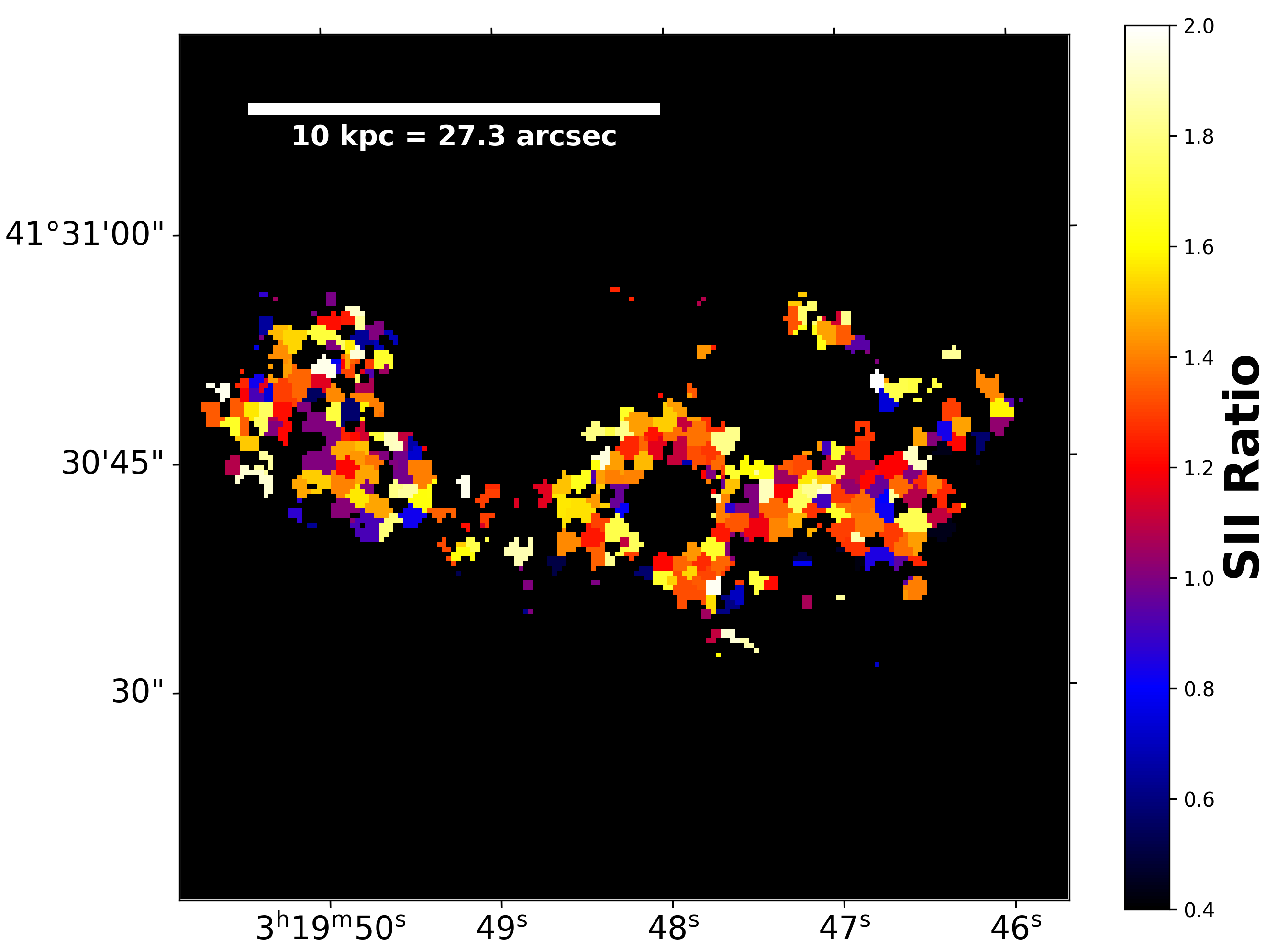}}
\hspace{-1.5cm}
\begin{minipage}[t]{0.47\textwidth}
  \usebox{\imagebox}
\end{minipage}
\hspace{0.75cm}
\begin{minipage}[t]{0.33\textwidth}
  \centering\raisebox{\dimexpr\ht\imagebox-\height}{\includegraphics[width=\textwidth]{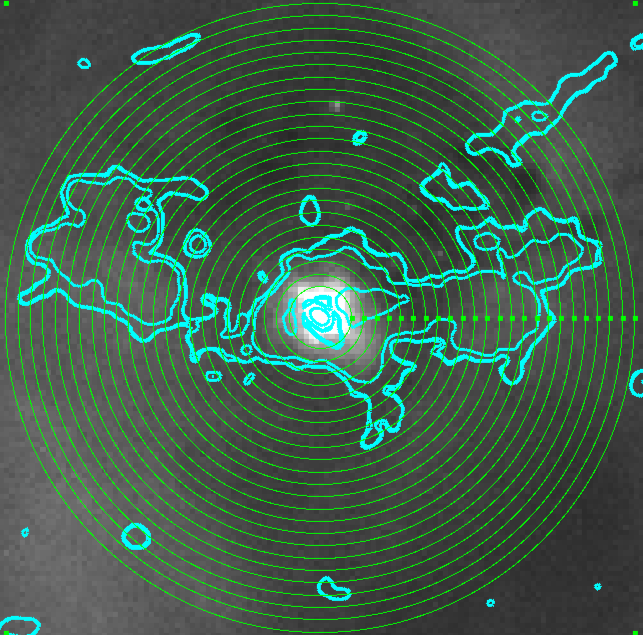}}
\end{minipage}
\begin{minipage}[t]{0.49\textwidth}
  \centering
  \includegraphics[width=\textwidth]{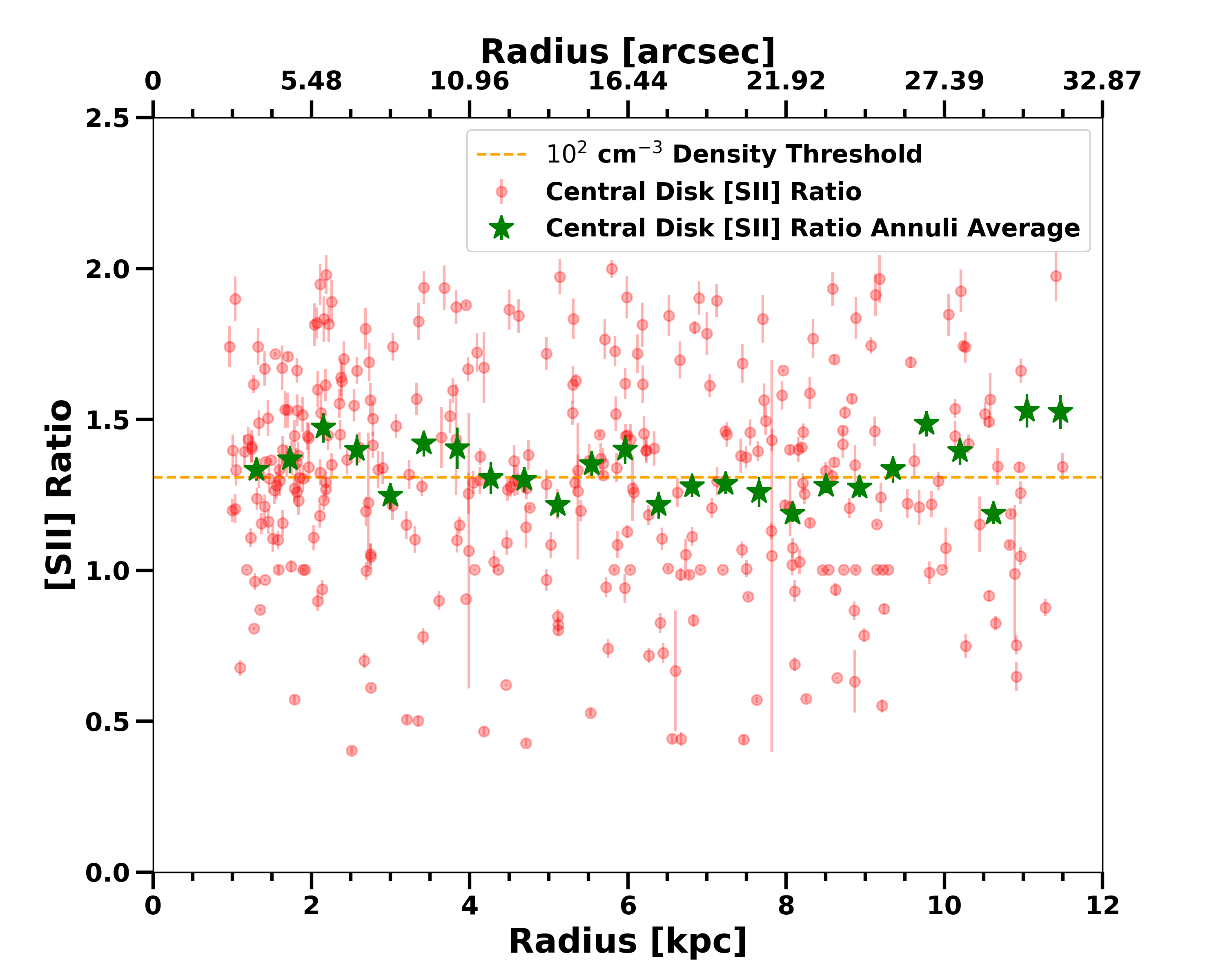}
\end{minipage}
\begin{minipage}[t]{0.49\textwidth}
  \centering
  \includegraphics[width=\textwidth]{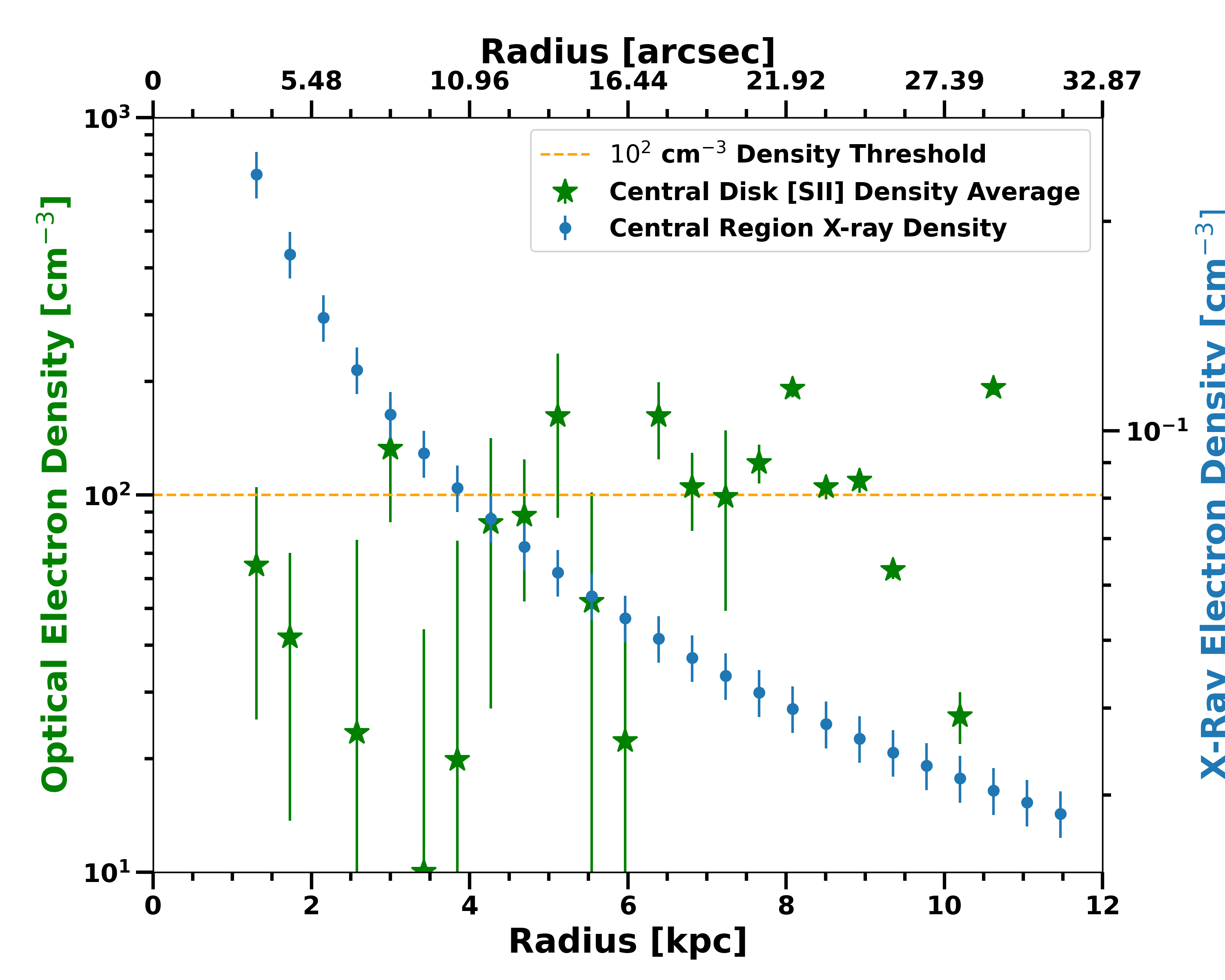}
\end{minipage}
\caption{Top Left: \sii\text{} ratio map of the central region obtained from the WVT algorithm. We divided here the resulting \sii$\lambda6716$ and \sii$\lambda6731$ flux maps, where we only consider the SITELLE spectra displaying a flux higher than $1\times10^{-17} \text{ erg s}^{-1}\text{cm}^{-2}\text{\AA}^{-1}$ tracing the central structure (\citealt{2024ApJ...962...96V}). Top Right: Chandra 0.5 - 2.0 keV observations of NGC 1275. The cyan contours show the H$\alpha$ filaments as observed with SITELLE at high-spectral resolution at a flux level between $1\times10^{-17} \text{ erg s}^{-1}\text{cm}^{-2}\text{\AA}^{-1}$ and $1\times10^{-16} \text{ erg s}^{-1}\text{cm}^{-2}\text{\AA}^{-1}$ for illustration purposes  (\citealt{2024ApJ...962...96V}). The 25 green annuli represent the regions considered to determine the X-ray electron density of the surrounding ICM. Bottom Left: Radial profile of the \sii\text{} emission line doublet ratio as a function of the distance from the center for all Voronoi cells as shown in the upper left panel in red data points, while the green stars indicate the average \sii\text{} ratio obtained within 25 annuli centered on the central galaxy. Their corresponding error bars represent the standard deviation of the measured \sii\text{} emission line ratios within the 25 annuli. The \sii\text{} ratios displayed here are only those found to be in the interval of 0.4 to 2.0. The dotted orange line corresponds to the low density limit of $n_e = 10^{2}\text{ cm}^{-3}$. Bottom Right: Radial profile of the derived optical electron density represented with green stars, obtained from the average \sii\text{} ratio of the bottom left panel using Equation \ref{eq:density_proxauf} (\citealt{proxauf2014upgrading}), and the X-ray electron density in blue data points as a function of the distance from the central galaxy. Due to a discontinuity in Equation \ref{eq:density_proxauf}, ratios above $\gtrsim 1.49$ cannot be used to derive a proper electron density. Four average ratios data points above this threshold value are therefore not considered in our calculations. The dotted orange line corresponds to the low density limit of $n_e = 10^{2}\text{ cm}^{-3}$.}
\label{fig:maps_central_structure}
\end{figure*}

Similarly, we can compare the distributions as a function of the flux normalized by the number of pixels belonging to each region. This analysis is displayed in the left panel of Figure \ref{fig:flux_distribution}, where data points are color-coded in a similar fashion as Figure \ref{fig:ratio_distribution}, and a similar dichotomy between these region seems to appear where the overall \sii\text{ }flux and ratio of the outer filaments are low ($\sim 2.5\times10^{-19} \text{ erg s}^{-1}\text{cm}^{-2}\text{\AA}^{-1}$ and $\sim 1.1$ on average respectively), whereas the central bright and turbulent region indeed showcases an average higher flux ($\sim 6\times10^{-18} \text{ erg s}^{-1}\text{cm}^{-2}\text{\AA}^{-1}$), but also higher average ratios ($\sim 1.34$) than the outer filaments. When comparing the distribution of the \sii\text{ }ratio as a function of radius centered on the AGN, as seen in the right panel of Figure \ref{fig:flux_distribution}, we can also observe more clearly this dichotomy between central and outer filaments. Indeed, this panel demonstrates that at lower radius close to the AGN below $\sim 10$ kpc, the \sii\text{ }ratio clusters around $\sim 1.34$, while for outer filaments at radius above $\geq 10$ kpc, the \sii\text{ }ratio falls to around $\sim 1.1$ as discussed previously. For the outer filaments, the errorbars displayed are obtained by propagating the errors returned from the \texttt{LUCI} fitting procedure we described in Section \ref{sii_fitting}. For the central structure however, we display the mean values of \sii\text{} emission line ratios obtained using the annuli shown in the upper right panel of Figure \ref{fig:maps_central_structure}. The corresponding error bars represent the standard deviation of the data points as seen in the bottom left panel of Figure \ref{fig:maps_central_structure}.





\begin{figure*}
  \centering
  \includegraphics[width=\linewidth]{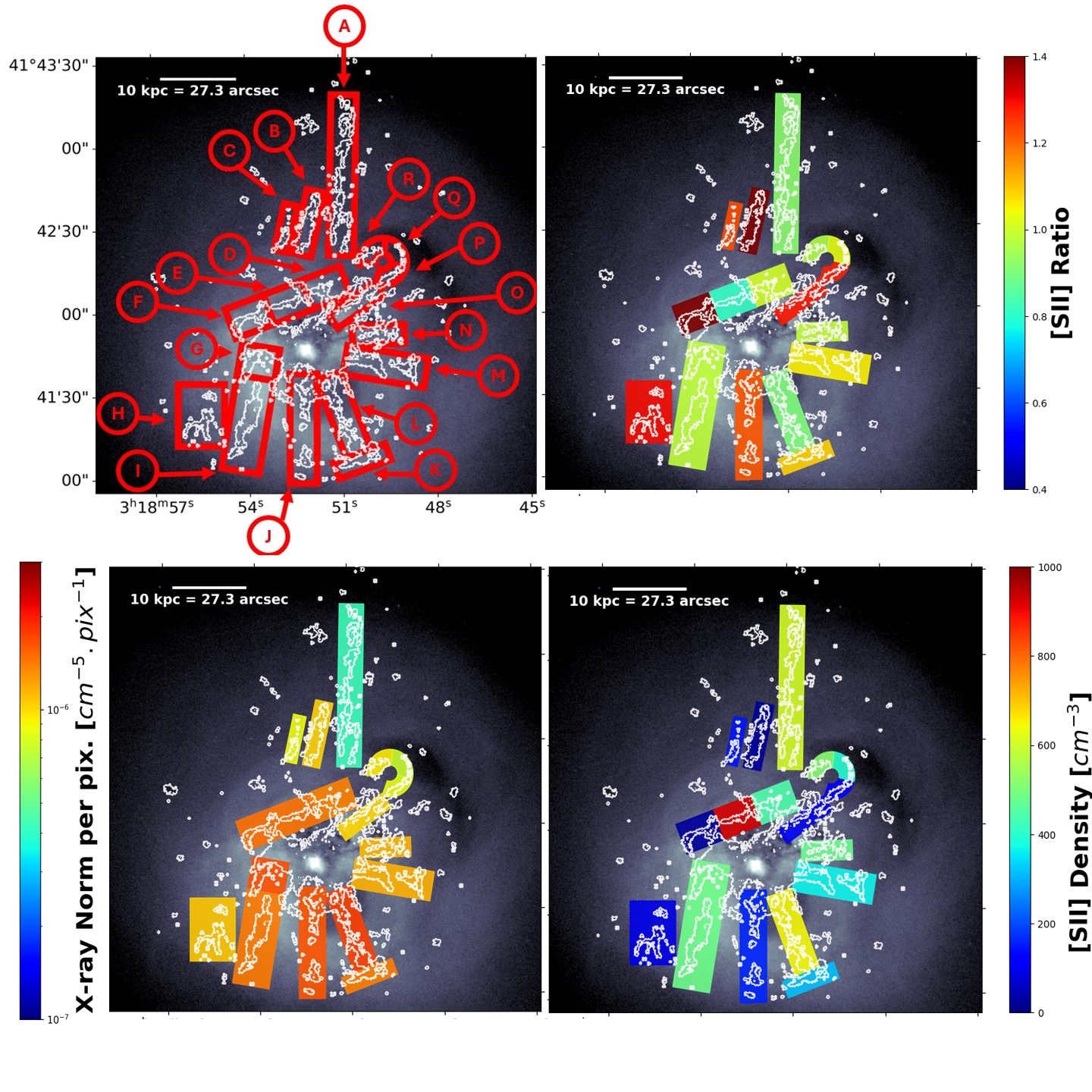}
\caption{Top Left: Map of the binned regions used for fitting the \sii\text{} emission line doublet ratio and corresponding electron density. The filament contours are displayed in white and overplotted on an X-ray greyscale image of the surrounding ICM emitting in X-ray. Top Right: \sii\text{} emission line doublet ratio map of the outer filament regions around NGC 1275. Bottom Left: X-ray norm divided by the regions area in pixels and derived from archival deep Chandra observations of the surrounding ICM obtained from the same regions used to determined the \sii\text{} ratio and electron densities of the outer filaments of NGC 1275. Bottom Right: electron density map of the outer filament regions around NGC 1275, as estimated through the \sii\text{} ratio relationship (see Section \ref{optical_electronic_density} and Equation \ref{eq:density_proxauf}).}
\label{fig:maps_outer_filaments}
\end{figure*}

\begin{deluxetable*}{l c c c c c}
\tablecaption{Results of the \sii\text{} emission-line doublet ratio and corresponding electron density for each binned region in the outer filaments. The first column gives the label used to identify each region in Figure~\ref{fig:maps_outer_filaments}. We indicate in the second column the distance in kpc between the geometrical center of each region and NGC 1275's center. The third column lists the \sii\text{} line ratio and its uncertainty, while the fourth column provides the corresponding electron density derived using the updated calibration of \citet{proxauf2014upgrading} (see Equation~\ref{eq:density_proxauf}). The fifth column reports the X-ray normalization per region area in pixels measured from archival \textit{Chandra} observations of the ICM, which serves as a proxy for the integral of the squared electron density along our line of sight. The sixth column gives the corresponding optical pressure, computed from the total particle density ($\sim 2n_e$; see Equation~\ref{eq:pressure}).
\label{tab:2}}
\tablehead{
\colhead{Region} &
\colhead{Distance from Center} &
\colhead{\sii\text{} Ratio} &
\colhead{Optical Density $n_e$} &
\colhead{X-ray Norm per pixel} &
\colhead{Optical Pressure $P$} \\
&
\colhead{(kpc)}&
&
\colhead{(cm$^{-3}$)} &
\colhead{(cm$^{-5}$ pix$^{-1}$)} &
\colhead{(erg cm$^{-3}$)}
}

\startdata
A & $35.3$ & $0.919\pm0.007$ & $607\pm16$ & $4.2\times10^{-7}$ & $1.68\pm0.04\times10^{-9}$ \\
B & $25.0$ & $1.398\pm0.060$ & $24\pm37$ & $1.0\times10^{-6}$ & $6.63\pm0.10\times10^{-11}$\\
C & $25.5$ & $1.231\pm0.348$ & $148\pm547$ & $8.8\times10^{-7}$ & $4.09\pm15.1\times10^{-10}$\\
D & $13.6$ & $1.003\pm0.013$ & $444\pm23$ & $1.4\times10^{-6}$ & $1.23\pm0.06\times10^{-9}$\\
E & $10.4$ & $0.811\pm0.023$ & $922\pm93$ & $1.5\times10^{-6}$ & $2.55\pm0.26\times10^{-9}$ \\
F & $13.3$ & $1.397\pm0.055$ & $24\pm34$ & $1.5\times10^{-6}$ & $6.63\pm9.39\times10^{-11}$\\
G & $10.3$ & $0.980\pm0.039$ & $484\pm76$ & $1.7\times10^{-6}$ & $1.34\pm0.21\times10^{-9}$\\
H & $23.8$ & $1.297\pm0.048$ & $90\pm40$ & $1.1\times10^{-6}$ & $2.49\pm1.10\times10^{-10}$\\
I & $19.3$ & $0.980\pm0.032$ & $484\pm61$ & $1.4\times10^{-6}$ & $1.34\pm0.17\times10^{-9}$\\
J & $14.6$ & $1.217\pm0.036$ & $161\pm38$ & $1.6\times10^{-6}$ & $4.45\pm1.05\times10^{-10}$\\
K & $24.4$ & $1.088\pm0.024$ & $315\pm34$ & $1.4\times10^{-6}$ & $8.70\pm0.94\times10^{-10}$\\
L & $14.8$ & $0.903\pm0.022$ & $646\pm55$ & $1.7\times10^{-6}$ & $1.78\pm0.15\times10^{-9}$\\
M & $15.3$ & $0.984\pm0.037$ & $477\pm71$ & $1.2\times10^{-6}$ & $1.32\pm0.20\times10^{-9}$\\
N & $13.2$ & $1.054\pm0.039$ & $364\pm61$ & $1.1\times10^{-6}$ & $1.01\pm0.17\times10^{-9}$\\
O & $12.8$ & $1.286\pm0.030$ & $99\pm25$ & $1.0\times10^{-6}$ & $2.73\pm0.69\times10^{-10}$\\
P & $21.1$ & $1.277\pm0.043$ & $107\pm38$ & $8.6\times10^{-7}$ & $2.95\pm1.05\times10^{-10}$\\
Q & $25.6$ & $1.031\pm0.491$ & $398\pm4.4\times10^3$ & $7.4\times10^{-7}$ & $1.10\pm12.1\times10^{-9}$\\
R & $22.2$ & $0.965\pm0.020$ & $512\pm39$ & $9.4\times10^{-7}$ & $1.4\pm0.11\times10^{-9}$\\
\enddata
\end{deluxetable*}

\subsection{Optical electron Density}
\label{optical_electronic_density}

Based on the \sii\text{} doublet ratio obtained previously, we are able to determine the electron density of the optically emitting ionised gas. Collisional deexcitation of metastable atomic energy levels can cause the level population to depend on gas density, allowing suitable line ratios to be used as a probe of the electron density. Spectroscopically, this analysis is undertaken by determining the flux ratio of two emission lines of the same ion with different energy levels but similar excitation energy. This allows the relative excitation rates of both energy levels to only be dependent on the ratio of collision strengths. Moreover, if we consider specific emission lines whose energy levels have different radiative transition probabilities or different collisional deexcitation rates, the corresponding ratio of lines intensities will inform us on the electron density of the gas. One of these line ratios is \sii$\lambda6716$ by \sii$\lambda6731$ (see \citealt{1989agna.book.....O}).

\cite{1989agna.book.....O} determined the density dependence of this line ratio, as well as lower and upper thresholds on the density range over which it can be used to determine the electron density. However, recent improvements to this formula have been developped by \cite{proxauf2014upgrading} to provide an effective and easy to use mathematical relationship to determine the electron density from the \sii\text{} emission lines ratios by using improved atomic data.

We therefore used the updated formula of \cite{proxauf2014upgrading} in order to obtain the electron density measurements of the ionised gas composing the filamentary nebula surrounding NGC 1275, which, for a gas temperature of $T=10000$ K, is given by:
\begin{equation}
    \begin{split}
        \text{log}(n_e[\text{cm}^{-3}]) = &\text{ } 0.0543 \text{ tan}(-3.0553\text{ }r + 2.8506)\\
        & + 6.98 - 10.6905\text{ }r\\
        & +9.9186\text{ }r^2 - 3.5442\text{ }r^3,
    \end{split}
    \label{eq:density_proxauf}
\end{equation} 
where $r$ corresponds to the measured ratio of \sii$\lambda6716$ and \sii$\lambda6731$. The resulting density map for the outer filaments can be seen in the bottom right panel of Figure \ref{fig:maps_outer_filaments}, while the corresponding values for electron densities of each regions are presented in Table \ref{tab:2}. We can see that the overall electron density is low at around $\sim 100$ to $500$ cm$^{-3}$ for most of the outer regions. Nevertheless, all of them fall in the expected interval between 0.4 to 1.32, corresponding to the \cite{1989agna.book.....O} and \cite{proxauf2014upgrading} boundaries of the \sii\text{ }emission line ratio, thus allowing us to obtain electron density measurements.

On the other hand, for the central bright and turbulent region seen in \cite{2024ApJ...962...96V}, ratios obtained for the \sii\text{ }emission line doublet are mostly at or above the low density threshold preventing us from determining meaningful values. These corresponding electron density values are shown in the bottom right part of Figure \ref{fig:maps_central_structure}, while the ratio map and \sii\text{ }ratio radial profile of the central region can be seen in the top left and bottom left panels of Figure \ref{fig:maps_central_structure} respectively. Similarly, we gathered the resulting ratio annuli measurements within the top histogram of Figure \ref{fig:ratio_distribution} where the corresponding electron density values have been added on the upper x-axis. As can be seen, the ratios obtained are mostly below the low-density threshold of $n_e = 10^{2}\text{ cm}^{-3}$ preventing us from drawing meaningful electron density results. Interestingly, we also do not seem to discern any radial trend with the ratios remaining globally high at around $\sim 1.34$ on average with a somewhat significant spread of $\sim 0.25$ and standard deviation of $\sim 0.098$.  We note however that an important scatter is visible for a few ratios obtained for the inner filaments, as visible for the red data points within the bottom left plot of Figure \ref{fig:maps_central_structure}. We suggest that this wide scatter could be due to the presence of unresolved multiple emission components found in the inner region of the filaments. Indeed, despite the relatively high spectral resolution of the SITELLE observations (R = 7000), we are only able to fit the observed spectra with a single emission component. However, archival ESPaDOnS spectro-polarimetric observations at a spectral resolution of R = 68000, showed the detection of multiple emission components in a $1.2$’’ wide region located $\sim 6$’’ away from the central galaxy, that were undetected within the SITELLE data (Vigneron et al. in preparation). This suggests that additional emission components could be present and not resolved in the observed spectra for the central region, which could lead to a significant scatter in the resulting \sii\text{} emission line ratios values of the observed summed emission components, each with their independent ratios. We will discuss the implications of our electron density results in Section \ref{section:discussion}.



\subsection{X-ray electron Density}
\label{X-ray_electron_density}

\subsubsection{Outer Regions}

As explained in Section 2.1.1, since the flux of the \sii\text{} lines was extremely low for the outer filaments, we decided to bin the pixels belonging to them into larger regions. The resulting regions are shown in the top left corner of Fig. \ref{fig:maps_outer_filaments}. They typically have sizes of several kpc/arcseconds and are spread throughout the cluster core. An issue with this technique is that we cannot deproject the spectra since deprojection normally assumes spherical geometry. Therefore, we must compare the \sii\text{} electron density to a projected proxy of the density for the X-ray emitting ICM. When fitting the X-ray spectra, one of the quantities extracted is called the norm and defined as
\begin{equation}
    \begin{split}
        \text{norm}= &\frac{10^{-14}}{4\pi[D_a(1 + z)]^2}\int n_e n_\mathrm{H} dV .
    \end{split}
\end{equation}
The norm depends on the redshift $z$, angular diameter distance $D_a$ and hydrogen density $n_\mathrm{H}$ as well as the electron density $n_e$. Given the shape of the regions used and since we are unable to determine the exact volume of the filaments, it is difficult to extract the density from the norm. Indeed, no formal analysis of their 3D structure has ever been attempted due to the extremely low velocity difference between emission components within the emission lines complex of the optical filaments. Instead, we choose to compare the \sii\text{} derived electron density variations to the X-ray norm variations divided by the regions pixel area, which is proportional to the average value over a region of the line integral of the electron density squared. These measurements are shown in the bottom left corner of Fig. \ref{fig:maps_outer_filaments}. We stress once more that our ability to perform a clear spatial correlation analysis between the optical and X-ray electron densities are hampered by the difficulty to obtain a volume measurement for the filaments. Nevertheless, these results interestingly show a dichotomy between the northern and southern regions. Indeed, we can see that the norm values divided by pixel area appear higher in the southern and eastern regions, while they are slightly lower in the northern filaments. The Horseshoe filament also appears to show a slightly lower norm per unit area  at around $\sim 7\times10^{-7} \text{cm}^{-5}\text{pix}^{-1}$.

\subsubsection{Central Region}

After fitting the X-ray emission spectra of the ICM, we are able to extract the radial density profile according to the annuli defined in the top right panel of Figure \ref{fig:maps_central_structure}. 
The electron density is determined through spectral deprojection by modeling the X-ray emission as a series of concentric spherical shells. Assuming spherical symmetry, the contribution of each shell to the projected line-of-sight emission is subtracted iteratively from the outside in, and the electron density is then derived from the deprojected emission measure and the volume of each shell.
To do so we considered the same annuli regions used to determine the \sii\text{ }emission line ratios for the central region.

As seen in the bottom right panel of Figure \ref{fig:maps_central_structure}, when specifically considering the density of the central annuli regions used to study the bright and turbulent structure described in \cite{2024ApJ...962...96V},  we clearly see a smooth profile showing a higher density of $\sim 2 \times 10^{-1} \text{ cm}^{-3}$ near the center, which then drops as a function of radius down to $<3 \times 10^{-2} \text{ cm}^{-3}$ at $11.5$''. This density profile is a clear characteristic of cool-core galaxy clusters where the ICM distribution is extremely peaked near the central BCG (see \citealt{churazov2003xmm}). 

\section{Discussion}
\label{section:discussion}

\begin{figure*}[h!]
  \centering
  \includegraphics[width=\linewidth]{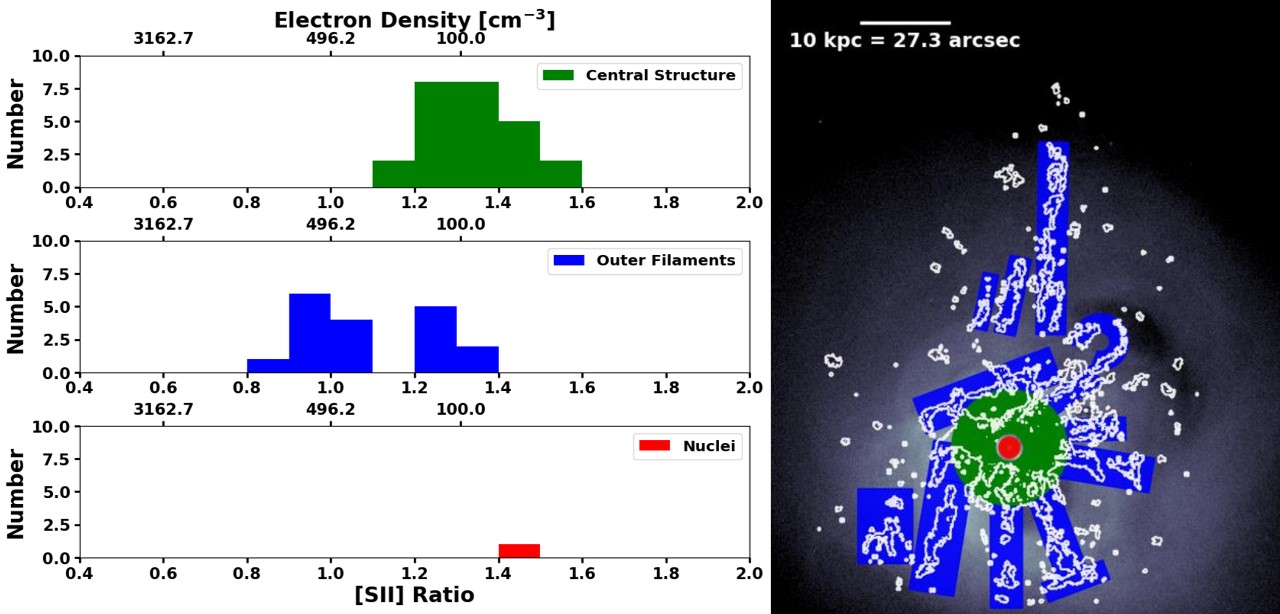}
\caption{Top Left: Distribution of the \sii\text{} emission line ratios for the central region of the filaments surrounding NGC 1275 as obtained from Figure \ref{fig:maps_central_structure}. Middle Left: Distribution of the \sii\text{} emission line ratios for the outer filaments surrounding NGC 1275 as obtained from Figure \ref{fig:maps_outer_filaments}. Bottom Left: Measurement of the \sii\text{} ratio from a central pointing encompassing the filaments and AGN emission profile of NGC 1275. Right: Color-coded map showing the locations of the central structure, outer filament and nucleus overplotted on white contours showing the optical filaments and greyscale background displaying the ICM X-ray emission.}
\label{fig:ratio_distribution}
\end{figure*}

\begin{figure*}
  \centering
  \includegraphics[width=\linewidth]{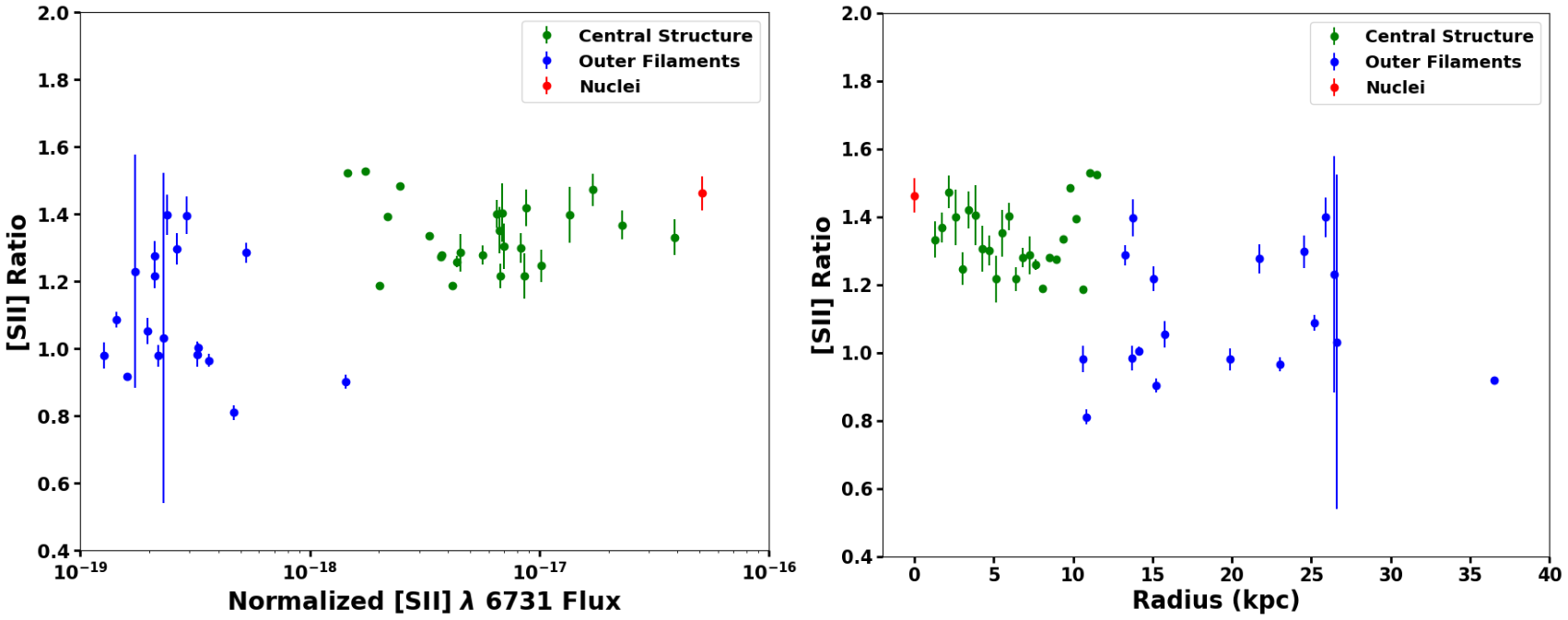}
\caption{Left: Distribution of the \sii\text{} emission lines ratios as a function of the normalized flux of the \sii$\lambda$6731 emission line. The data points are color-coded in a similar fashion as Figure \ref{fig:ratio_distribution} to differentiate their locations. Error bars are either derived from error propagation using \texttt{LUCI}'s fitting procedure for the outer filaments and nucleus or from the standard deviation of the distribution of \sii\text{} emission line ratio within the inner 25 annuli for the central structure. Right: Distribution of the \sii\text{} emission lines ratios as a function of radius from the central AGN. The data points are color-coded in a similar fashion as Figure \ref{fig:ratio_distribution} to differentiate their locations. Error bars are either derived from error propagation using \texttt{LUCI}'s fitting procedure for the outer filaments and the nucleus or from the standard deviation of the distribution of \sii\text{} emission line ratio within the inner 25 annuli for the central structure.}
\label{fig:flux_distribution}
\end{figure*}


\subsection{Density Comparison between Optical Filaments and ICM}


We first compare the results obtained in terms of electron density, \sii\text{ }emission line ratios and norm per unit area values of the X-ray spectral fitting between the SITELLE and Chandra observations to explore potential spatial correlations or differences.

\subsubsection{Central Structure}

Considering first the central emission from both the ICM and the central bright optical structure detailed in \cite{2024ApJ...962...96V}, we can see significant differences between their density structures. Indeed, the ICM displays the clear properties of a cool-core cluster where the electron density decreases steadily from the central galaxy, going from $n_e = 2 \times 10^{-1}\text{ cm}^{-3}$ to $n_e = 7 \times 10^{-3}\text{ cm}^{-3}$ at a radius of around 10 kpc as can be seen in the bottom right panel of Figure \ref{fig:maps_central_structure}. On the other hand, the central bright structure visible in the optically emitting filamentary nebula mostly displays a stable trend of high \sii\text{ }emission line ratios around $\sim 1.34$, which translates into a low density structure with little radial variations, mostly at or below the low density limit of $n_e = 10^2\text{ cm}^{-3}$. Therefore, no clear correlation between the two environments in terms of electron density can be drawn and we will explore in Section 4.2 potential models to explain the apparent radially stable density profile of the optical filaments.

\subsubsection{Outer Filaments}

Focusing now on the outer filaments regions, we can obtain clear electron density measurements in the optical, but we can only rely on the norm per unit area of the X-ray spectra fitting as proportional to the average value over a region of the line integral of the electron density squared, as explained in Section \ref{X-ray_electron_density}. It appears noteworthy that we do not seem to be able to see any clear correlations between the two maps presented in the bottom left and right panels of Figs. \ref{fig:maps_outer_filaments}. On one hand, the extended optical filaments seem to display a moderate density overall at around $n_e = 5 \times 10^2\text{ cm}^{-3}$ to $n_e = 10^2\text{ cm}^{-3}$. On the other hand, the X-ray norm divided by the regions pixel area displayed in the bottom left corner of Figure \ref{fig:maps_outer_filaments} seems to introduce a dichotomy between the southern and eastern regions displaying higher values ($\text{norm} \sim 10^{-6}\text{ cm}^{-5}\text{pix}^{-1}$) versus the northern ones where the norm per unit area is slightly lower ($\text{norm} \sim 7 \times 10^{-7}\text{ cm}^{-5}\text{pix}^{-1}$), likely resulting from projection effects given that the X-ray electron density roughly scales with $\sim 1/r$, where $r$ represents the distance from the cluster center. We indicate in the second column of Table \ref{tab:2} the distance in kpc between the center of each region and NGC 1275, reinforcing this interpretation as the most distant regions indicate a lower X-ray norm per unit area overall.

An interesting exception visible in the bottom right corner of Figure \ref{fig:maps_outer_filaments} is the single region denoted as E and located slightly above the northern X-ray cavity, which stretches over 8.9 by 6.4 kpc, displaying a high density of $n_e = 922 \pm 93\text{ cm}^{-3}$. The position of this specific region and its corresponding high electron density could be the result of compression of the ionised gas by the currently-forming radio bubble. However, since we do not possess a clear three-dimensional representation of the filamentary nebula, we cannot say with certainty that this eastern filament is located directly above the radio bubble. Nevertheless, it is interesting that this region shows such a high electron density compared to the rest of the outer filaments. The corresponding X-ray norm per unit area for this region, as seen in the bottom left corner of Figure \ref{fig:maps_outer_filaments}, does not appear significantly different than the other surrounding filaments and shows a value of norm per unit area of $\sim 10^{-6}\text{ cm}^{-5}\text{pix}^{-1}$. A higher norm per unit area can however be observed for the southern filament regions crossing over the southern ghost cavity where we observe values of norm per unit area $\sim 2 \times 10^{-6}\text{ cm}^{-5}\text{pix}^{-1}$, which could indicate a trend of slightly elevated electron density in both media crossing over the southern ghost cavity.

Interestingly, the Horseshoe filament also appears to show similar trends in both optical and X-rays. The bottom left corner of Figure \ref{fig:maps_outer_filaments} clearly indicates a lower X-ray norm per unit area at around $\sim 7\times 10^{-7}\text{ cm}^{-5}\text{pix}^{-1}$ compared to the other regions, which could be the result of projection effects as the Horseshoe filament is located quite further away from the cluster center (see second column of Table \ref{tab:2}), thus impacting the derived X-ray norm per unit area as mentioned previously. Similarly, the optical density map in the bottom right panel of Figure \ref{fig:maps_outer_filaments} shows a relatively low density at $n_e = 10^2\text{ cm}^{-3}$, increasing slightly to $n_e = 4 \times 10^2\text{ cm}^{-3}$ below the cavity observed in X-ray. However, as indicated in Table \ref{tab:2}, the electron density error for the top of the Horseshoe filament is extremely high, as the SITELLE observations only detect a limited number of pixels for this part of the filament, the \sii\text{} ratio error appears relatively high leading to a high density error overall. Nevertheless, this issue is not observed for the other two regions covering the Horseshoe filament. We hypothesize that the presence of the X-ray ghost cavity directly above the Horseshoe filament could influence the observed lower X-ray norm per unit area and lead to the increasing optical density behind it. 


Finally, the south-eastern region of the filaments called the 'Blue Loop' (see Fig. \ref{fig:fov}), due to its high concentration of younger star clusters (\citealt{hatch_origin_2006}), does not indicate an increased optical electron density and instead shows lower values around $n_e \sim 1 \text{ to } 5 \times 10^2\text{ cm}^{-3}$ as displayed in the bottom right panel of Figure \ref{fig:maps_outer_filaments}. Interestingly, the X-ray norm per unit area displays a similar trend compared to the optical electron density between the two regions covering the Blue Loop, one of them remaining slightly lower overall at $\sim 1 \times 10^{-6}\text{ cm}^{-5}\text{pix}^{-1}$, while the other displays a slightly higher norm per unit area  at $\sim 1.5 \times 10^{-6}\text{ cm}^{-5}\text{pix}^{-1}$, as can be seen in the bottom left panel of Figure \ref{fig:maps_outer_filaments}.

Nevertheless, in regard to the outer filaments' electron density measurements, since we considered large swathes of optical and X-ray gas of various dimensions to get a relevant electron density measurement, we must remain cautious in our comparison between the densities of the optically emitting cooler filaments and the hot X-ray emitting ICM. Indeed, smaller scale comparisons like the one proposed here for the Horseshoe filament can lead to tentative density trends between the two medium. Thus, obtaining deeper observations of the \sii\text{ }emission doublet across the entire filamentary nebula with a higher signal-to-noise ratio could prove fruitful to better determine if other density trends between the optical filaments and ICM could be observed on much smaller scales. Moreover, obtaining clearer density profiles across the outer filaments by using other emission line doublet and auroral lines could also help us determine radial trends as a function of distance from the central AGN, therefore allowing us to observe if a significant energetic contribution from the central SMBH can affect the emission profile of distant filaments. The inner region on the other hand displays a much higher \sii\text{ }signal which allows us to radially study the resulting electron density on a much smaller scale. 

As we have discussed in this section and seen in Figure \ref{fig:maps_outer_filaments} and \ref{fig:maps_central_structure}, it appears that no clear correlations in terms of electron density can be derived between the optically-emitting cool filaments and the X-ray emitting hot ICM based on the SITELLE and Chandra observations. Thus, we will first explore in the following subsection the relationship between density, temperature and pressure within the inner $\sim 10$ kpc of the multiphase environment surrounding NGC 1275 and discuss afterwards potential density models that could explain the results observed for the optical filaments.

\subsection{Relationship Between Density, Temperature and Pressure}
\label{density_vs_temp}

\begin{figure*}
  \centering
  \includegraphics[width=0.95\linewidth]{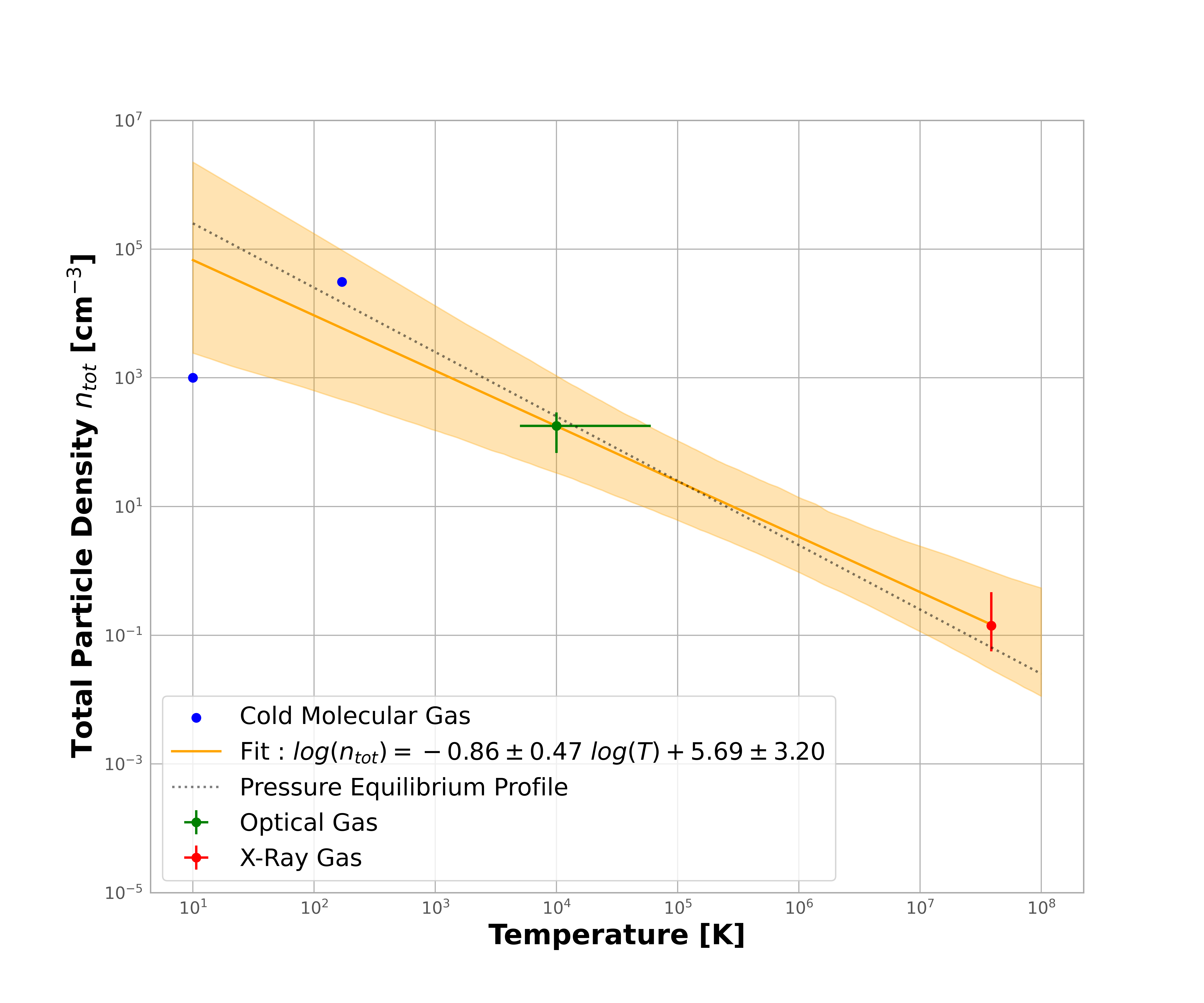}
\caption{Total particle density as a function of temperature for the cold molecular gas (blue points) derived from \cite{bridges1998molecular}, averaged optical filaments with SITELLE and averaged X-ray emitting ICM with Chandra within the central $\sim 10$ kpc from NGC 1275. The best linear fit is displayed in orange with the shaded area corresponding to the 3$\sigma$ error on the fit. A pressure equilibrium profile with a slope of $-1$ has also been included with a grey dotted line over the current fit.}
\label{fig:density_vs_temp}
\end{figure*}

Following our spatial analysis comparing the electron density between the optical filaments and the X-ray emitting ICM, we also performed a minor study including measurements from cold molecular gas found within the central $21$'' around NGC 1275 (\citealt{bridges1998molecular}). In order to determine a relationship between these three phases, we decided to plot the measurements of density as a function of temperature for each by taking the two temperature and density model for the cold molecular gas as derived by \cite{bridges1998molecular} (namely a 10 K cold component with a density of $n_{mol H} = 10^3 \text{ cm}^{-3}$ and a 170 K warmer component with a density of $n_{mol H} = 3.1\times10^4 \text{ cm}^{-3}$), while taking the averages of total density and temperature for the optical and X-ray measurements as obtained with the SITELLE and archival Chandra observations in the central structure near the BCG and described in Sections 3.2 and 3.3 (respectively $10^4$ K and $n_{tot} = 178.6 \text{ cm}^{-3}$ for the optical filaments and $3.9 \times 10^7$ K and $n_{tot} = 1.4 \times 10^{-1} \text{ cm}^{-3}$ for the X-ray emitting ICM). To accurately compare the density and pressure measurements between the cold molecular gas and ionised gas phases, we consider here the total density of the ICM and optical filaments by assuming that $n_{tot} \sim 2n_e$, since the number of particles per electron would be closer to 2. When reporting these measurements in a log-log space, a linear trend can be determined, the slope of which would inform us on the relative pressure profile between the three phases, as can be seen in Figure \ref{fig:density_vs_temp}. This, in turn, could inform us on the potential equilibrium or disequilibrium that could occur in the complex multiphase environment surrounding NGC 1275.

Therefore, we fit a linear relationship between the four points obtained in Figure \ref{fig:density_vs_temp} using a MCMC approach implemented with the \texttt{emcee} package, where we used 2000 walkers and a likelihood function incorporating errors on both temperature and density. The resulting posterior distributions thus yield the most probable slope and intercept. 
As can be seen in Figure \ref{fig:density_vs_temp}, our best fit, as obtained with the MCMC algorithm, is $\text{log}(n_{tot}) = (-0.86 \pm 0.47) \text{log}(T) + 5.69\pm 3.20$. Interestingly, the overall slope of our linear fit is slightly greater than but close to minus one thus indicating that pressure equilibrium exists broadly between the three phases. Indeed, in the context of pressure equilibrium, we would expect $n_{tot} k_\mathrm{B}T = P$, where $P$ would simply be a constant and $k_\mathrm{B}$ is Boltzmann's constant. Thus rearranging our terms and taking the base 10 logarithm would give us that $n_{tot} = P/ k_\mathrm{B}T$ becomes $\text{log}(n_{tot}) = -\text{log}(T) + \text{constant}$. 

Therefore, this simple fitting trend appears to indicate tentative signs of pressure equilibrium between the various phases studied here. We note, however, that such pressure equilibrium would be expected as gas phases in physical contact over time would naturally tend to come to equilibrium. The different phases may nevertheless have formed in distinct regions of the cluster atmosphere and at different pressures than those currently observed. 

Exploring the broad pressure equilibrium between all three phases, we determined approximate pressure values for each using our average density and temperature measurements. Indeed, for the cold molecular gas, we can assume a pressure relationship following, $P = n_{mol} k_\mathrm{B}T$. For the ionised gas however, the number of particles per electron would be multiplied by a factor of 2 approximately, leading us to a pressure expression of :

\begin{equation}
    P \sim 2n_e k_\mathrm{B}T,
    \label{eq:pressure}
\end{equation}
which we use in Figure \ref{fig:density_vs_temp} for the inner filaments and ICM, as well as in Table \ref{tab:2} for the outer filaments. Consequently, the pressure derived for the cold molecular gas based on \cite{bridges1998molecular} results, is of $P_{mol}\sim 1.4\times 10^{-12}$ erg cm$^{-3}$ and $P_{mol}\sim 7.3\times 10^{-10}$ erg cm$^{-3}$ respectively for both emission components. On the other hand, the inner optical filaments show an average pressure of $P_{fil}\sim 2.5\times 10^{-10}$ erg cm$^{-3}$, while the surrounding ICM possesses an average pressure of $P_{ICM}\sim 7.6\times 10^{-10}$  erg cm$^{-3}$ within 10 kpc. We can therefore observe a tentative pressure equilibrium between all three phases in the inner $\sim 10$ kpc, when omitting the coldest molecular gas component. We also determined the associated pressure for each outer region of the optical filaments following Equation \ref{eq:pressure} and reported them in the last column of Table \ref{tab:2}. From these results, we can observe that the relative pressure between each region varies by a factor of $\sim 38$ (from $6.63\times10^{-11}$ to $ 2.55\times10^{-9}$ erg cm$^{-3}$), most likely due to the large swathe of filaments considered to determine the electron density, tentatively biasing the results toward higher density clumps. Interestingly, ICM pressure measurements from deprojected maps show similar pressure order of magnitudes measurements at around $\sim 10^{-10}$ erg cm$^{-3}$ for these regions up to around $\sim 40$ kpc away from the central galaxy (see \citealt{sanders2004mapping}, \citealt{fabian2006very}). Our results therefore appear to indicate that a broad pressure equilibrium exists between the optical filaments and ICM. However, additional observations at a much higher spectral resolution and exposure time would be necessary to reveal potential multiple emission components and clearer \sii\text{} emission lines in the outer filaments to perform a dedicated spatial analysis of the electron density and pressure within the filaments of ionised gas surrounding NGC 1275.

On the other hand, the relatively large error profile of our fit leaves room for a potential increase in non-thermal pressure in the cooler phases, especially if we apply more weight on the coldest molecular gas density measurement of \cite{bridges1998molecular}. This non-thermal pressure support could take the form of increased magnetic pressure allowing the gas to cool without the need to further increase its density. In the following subsection, we will thus explore possible density models for the optical filaments and surrounding ICM based on our spectroscopic observations and the dependence between density and temperature studied here.

\subsection{Density Models of the Optical Filaments}

Considering the significant dichotomy between the electron densities and \sii\text{} emission line doublet ratio of the central bright and turbulent structure compared to the outer and quiescent filaments as seen in Figs. \ref{fig:ratio_distribution} and \ref{fig:flux_distribution}, we investigate here potential models to explain these results. 

As we have explored in Section \ref{density_vs_temp} through a joint study of the pressure profiles within the inner $\sim 10$ kpc of the X-ray emitting ICM, optical filaments and cold molecular gas, our results are consistent with broad pressure equilibrium between these three phases. 
Although, the large error bars in our fitting trend could tentatively indicate an increase in non-thermal pressure affecting the inner multiphase environment around NGC 1275. 

Thus, in this simplified picture and considering the optical phase, we would expect the filament’s densities to decrease with radius, but as we have seen in the bottom panels of Figure \ref{fig:maps_central_structure}, their overall densities much rather appear to be constant as a function of radius. Therefore, this could support the idea that either non-thermal pressure is indeed increased in the warm and cold phases allowing the gas to cool without the need of increased density, or that mechanisms in place within the system could lead to an increase in density at larger radii in comparison to smaller radii, such as shocks for instance. We will thus discuss these possibilities in this subsection.

First, magnetic pressure has been demonstrated as a possible support mechanism for the formation and survival of optical filaments around BCGs of cool-core clusters (\citealt{fabian_magnetic_2008}). Therefore, an assumed increase in non-thermal pressure for the warm and cold phases could potentially happen through magnetic pressure support. Indeed, if the hot and weakly magnetized ICM cools radiatively, then magnetic fields would be adiabatically compressed and would dominate for cooler phases. As suggested in \cite{fabian_magnetic_2008}, increasing the density by a factor of $10^4$ (from $0.01$ to $100 \text{ cm}^{-3}$ as we have seen in the bottom right panel of Figure \ref{optical_electronic_density} between the X-ray and optical phases), under the assumption of isotropic compression under flux freezing,  should increase the magnetic field as: 

\begin{equation} 
 	B \sim B_0 \left(\frac{n_{\text{optical}}}{n_{\text{x-ray}}}\right)^{2/3} \sim 400 B_0,
    \label{eq:mag_field}
\end{equation} 

where $B$ is the optical filaments magnetic field strength, $B_0$ is the ICM magnetic field strength while $n_{\text{optical}}$ and $n_{\text{x-ray}}$ are their respective electron densities.
Therefore, if we consider the estimated ICM magnetic fields of $\sim 2.0$ $\mu$G (see \citealt{molendi2004intra, mernier2023discovery}), this would lead to a dominant magnetic field of up to $\sim 800$ $\mu$G for the filaments, much greater than the estimated value of $\sim 27$ $\mu$G needed to support the filaments (\citealt{fabian_magnetic_2008}), as well as derived magnetic fields of $\sim 30-70$ $\mu$G within M87's filaments (\citealt{werner2013nature}). The scaling $B \propto n^{2/3} $ presented in Equation \ref{eq:mag_field} assumes isotropic compression under flux freezing in the framework of ideal magnetohydrodynamics. This relation holds when the magnetic field is dynamically subdominant (ie. $P_B \ll P_{thermal}$) and the compression remains approximately isotropic. However, as the magnetic pressure approaches or exceeds the thermal pressure, magnetic stresses become anisotropic, preferentially resisting compression perpendicular to the field lines. In this regime, the amplification of the magnetic field is reduced and becomes dependent on the geometry of the flow, potentially deviating significantly from a $n^{2/3}$ scaling. Therefore, the estimated magnetic field strengths derived above should be regarded as upper limits. Nevertheless, even accounting for this limitation, magnetic fields at the level of a few tens of µG are sufficient to provide significant non-thermal pressure support, consistent with observational estimates in systems such as M87, mentioned previously, or NGC 4696 (see \citealt{fabian_magnetic_2008}, \citealt{fabian2016hst}, \citealt{werner2013nature}).
Consequently, if the optical filaments form out of the hot ICM through thermal instabilities at a larger radius of $\sim 10$ kpc, the resulting magnetic field amplification would render them magnetically dominated. In this regime, magnetic pressure support would allow the filaments to fall inward toward the cluster center without undergoing significant compression, and thus without a corresponding increase in electron density. 
This mechanism could tentatively explain why the filament's densities in the central $\sim 10$ kpc appear rather homogeneous at the level of $n_e \sim 100 \text{ cm}^{-3}$ (see bottom right panel of Figure \ref{optical_electronic_density}). Moreover, the observed diversity of optical electron densities throughout the outer filamentary nebula of NGC 1275 could also be due to varying magnetic pressure support.

Finally, another possible explanation for the observed \sii\text{ }emission lines features and derived electron densities could be the result of smearing from multiple emission lines components in a region where it would be more likely for filaments to overlap along the line-of-sight. Despite the high spectral resolution of the SITELLE observations, able to probe a velocity scale of around $\sim 40\text{ km s}^{-1}$, we are not able to disentangle multiple velocity components within the central region allowing us to confirm the presence of an overlap of filaments. Higher spectral resolution spectroscopy of the filaments surrounding NGC 1275 with better signal-to-noise ratios would allow us to detect multiple emission components and determine if overlapping structures can be separated. A dedicated analysis of archival CFHT/ESPaDOnS observations of several pointings within the filamentary nebula surrounding NGC 1275 at an extremely high spectral resolution (R = 68000) is currently underway and could help us better understand the optical emission behavior and structure (Vigneron et al., in preparation).


\subsection{Other Observations of Filamentary Nebula around BCGs}

As we previously mentioned, multiple observations of optical filamentary nebula surrounding BCGs of galaxy clusters have been carried out and studies of their \sii\text{ }emission line doublet have shown that ratios often lie close to the lower density limit of the \cite{1989agna.book.....O} relationship (see for instance \citealt{farage2010optical}, \citealt{mcdonald2012optical}, \citealt{hamer_optical_2016}, \citealt{iani2019inquiring}, \citealt{ciocan2021vlt}, \citealt{tamhane2023radio}). As did previous spectroscopic studies of the \sii\text{ }emission doublet of the filamentary nebula surrounding NGC 1275 (\citealt{heckman1989dynamical}, \citealt{sabra2000emission} and \citealt{hatch_origin_2006}), we also confirm that a similar trend can be observed with new high spectral resolution SITELLE observations.

Therefore, a lower electron density in the optical filaments of ionised gas found around BCGs of cool-core galaxy clusters almost appears as a feature of these structures. The complex interplay between the AGN activity and the resulting \sii\text{ }emission line doublet ratio makes the use of the classical \cite{1989agna.book.....O} relationship for determining electron density much more difficult, since this often leads to values outside of the emission line ratio range needed to derive meaningful density measurements. As we have explored in Section 4.2, magnetic pressure support in the inner region could prevent an increase of density in the warm optical phase thus affecting the observed \sii\text{ }emission line ratio.

\cite{olivares2025halpha} performed a detailed analysis of the temperature, pressure and electron density of both optical and X-ray filamentary structures around the BCGs of the Centaurus cluster and M87. They discovered a clear surface brightness correlation between these filaments suggesting a strong connection between these two gas phases and a possibly similar ionizing mechanism. 
These results also indicates low densities of $n_e = 60$ and $120 \text{ cm}^{-3}$ respectively for the optical filaments at a temperature of $T = 10000$ K. The X-ray filaments on the other hand indicate a lower density of around $n_e = 0.02$ and $0.45 \text{ cm}^{-3}$. These values appear on the same order of magnitude as what we observe in the Perseus Cluster but we note that the authors used auroral lines and doublet of \nii\text{} in addition to the \sii\text{} emission line doublet to ascertain more precise and reliable density measurements. This allowed them to work around the limits of the \sii\text{} emission line ratio relationship to determine their corresponding electron densities.

It is therefore worth considering the innate difficulties of analyzing the \sii\text{ }emission line doublet. Indeed, in \cite{hamer_optical_2016} and \cite{tamhane2023radio}, the authors explained the complexity to retrieve correct flux measurements due to background emission and absorption effects. A similar issue regarding the \sii\text{ }emission lines within the filaments surrounding NGC 1275 is also present in the SITELLE observations, mainly for the outer filaments where the signal is lower overall and where background sky emission also severely affects our ability to retrieve the \sii\text{ }doublet to obtain spatially resolved electron density measurements on small scale. 

Interestingly, aside from \cite{olivares2025halpha}, no comparison between the electron density of the X-ray emitting environment in galaxy clusters and the optical filaments around their BCGs seems to have been derived in other cool-core galaxy clusters. An avenue of studies regarding the correlations in densities between these structures, as well as the cold molecular gas clouds and clumps in other systems could offer a better perspective regarding the potential similar trends in densities between these structures to what we hypothesized for the Horseshoe filament around NGC 1275. Nevertheless, a detailed study of the optical electron density in filamentary nebula surrounding BCGs requires a sufficiently high signal-to-noise to effectively carry out ratio measurements of the \sii\text{ }emission lines on small scales across filaments. However, as \cite{olivares2025halpha} have shown, a study of diverse density-sensitive emission line doublet as well as their auroral lines could offer better constraints on the electron density measurements performed across filamentary structures surrounding BCGs. 


\section{Conclusion} \label{sec:conclusion}

We analyzed new high-spectral resolution observations with SITELLE and determined the flux ratio of the emission lines \sii$\lambda6716$ by \sii$\lambda6731$ across the filamentary nebula surrounding NGC 1275, the brightest cluster galaxy of the Perseus cluster.
\begin{itemize}
    
    \item We determined the electron density of the entirety of the optically emitting filaments. These measurements indicate  that the ionised gas mostly lies  close to the low-density threshold of $n_e = 10^2\text{ cm}^{-3}$ across almost all the filaments. Interesting tentative trends such as increase or decrease in density can be observed and potentially be associated with the currently forming radio bubbles as well as older ghost cavities visible in the surrounding ICM.
    
    \item We observe a clear dichotomy in electron density between the inner filaments close to the central galaxy and the outer filaments. Similar differences were observed between these two regions in terms of flux and velocity dispersion as explored in \citealt{2024ApJ...962...96V}. The central, turbulent and luminous filaments display a higher ($\sim 1.3$) but seemingly radially constant \sii\text{ }emission line ratio while the outer, more quiescent and faint, filaments indicate a slightly lower \sii\text{ }emission line ratio of $\sim 1.1$.
    
    \item We compared the electron density of the optically emitting cooler filaments and the hotter X-ray emitting ICM to determine if any correlations could be found. Interestingly, some trends can be observed associated with specific filaments such as the Horseshoe filament and the eastern filament above the currently forming northern radio bubble. However, no clear correlation seems to be detectable overall in the extended filaments. Radial analysis close to the central galaxy indicates that the optical filaments' density appears relatively stable, while the surrounding ICM density is peaked towards the center and declines steadily as a function of radius. Considering the outer filament  locations, the associated X-ray norm divided by pixel area, used as a as a proxy for the integral along our line of sight of the squared electron density, indicates denser southern regions as opposed to less dense northern ones, likely due to projection effects. This dichotomy is, however, not detectable within the optical filaments' density profiles.

    \item Investigating the total particle density versus temperature profile between the cold molecular gas, optical filaments and X-ray emitting ICM phases through an MCMC fitting, leads to a power-law profile of $\text{log}(n_{tot}) = (-0.86 \pm 0.47) \text{log}(T) + 5.69\pm 3.20$. This result indicates tentative signs of pressure equilibrium between all three phases. The large error on the fit could also reveal the presence of an increased non-thermal pressure in the warm and cold phases, possibly due to magnetic fields. We also explore pressure measurements derived from our optical observations and X-ray deprojected pressure maps for outer filaments. A broad pressure equilibrium is also observed, although the large swathes of filaments considered in the optical analysis likely bias the results toward higher density clumps.
    
    \item We explored a potential density model for the optical filaments surrounding NGC 1275 involving the presence of significant magnetic pressure likely alleviating the need of a higher density in the central region close to NGC 1275.
\end{itemize}

Through our analysis of new high-spectral resolution observations of the filamentary nebula surrounding NGC 1275, we reinforced the previous electron density measurements established by past observations \citep{heckman1989dynamical,sabra2000emission,hatch_origin_2006} and expanded their scope to the entirety of the filaments. Such high spectral resolution observations of the filamentary nebula surrounding NGC 1275 allowed us to obtain new insights on the ionization mechanisms of the cooler optically emitting gas and its interactions with the warmer ICM in which it is embedded. Nevertheless, forthcoming X-ray observations of the Perseus Cluster with the XRISM space telescope (XRISM Science \citealt{team2020science}), successor of Hitomi, will enable a breakthrough in the study of AGN feedback. The analysis of SITELLE observations at high spectral resolution performed here will thus offer a detailed study of the electron density of the optical filaments for future X-ray density analysis of NGC 1275.\\


The authors would like to thank the Canada-France-Hawaii Telescope (CFHT) which is operated by the National Research Council (NRC) of Canada, the Institut National des Sciences de l'Univers of the Centre National de la Recherche Scientifique (CNRS) of France, and the University of Hawaii. Observations at CFHT were performed with care and respect from the summit of Maunakea which is a significant cultural and historic site. We also thank an anonymous referee for their constructive suggestions.

B.V. acknowledges financial support from the physics departement of the Université de Montréal. J.H.L. acknowledges funding support from the Canada Research Chairs Program, as well as the Natural Sciences and Engineering Research Council of Canada (NSERC) through the Discovery Grant and Accelerator Supplement programs.
C.P. acknowledges support by the European Research Council under ERC-AdG grant PICOGAL-101019746.
L.R.-N. is grateful to the National Science foundation NSF - 2109124 and the Natural Sciences and Engineering Research Council of Canada NSERC - RGPIN-2023-03487 for their support. N.W. acknowledges support by the GACR EXPRO grant 21-13491X.

B.V. also personally acknowledges Pr. Christopher Conselice for sharing observational data of NGC 1275 in the optical.


\software{python (\citealt{van2009python}), astropy (\citealt{robitaille_astropy_2013}, \citealt{price-whelan_astropy_2018}), numpy (\citealt{harris_array_2020}), scipy (\citealt{virtanen_scipy_2020}), matplotlib (\citealt{hunter_matplotlib_2007}), \texttt{LUCI} (\citealt{rhea_luci_2021})}, \texttt{Pumpkin} (\citealt{rhea_novel_2020}), sherpa (\citealt{2024ApJS..274...43S}), emcee (\citealt{2013PASP..125..306F}).

\clearpage
\appendix
\section{[SII] Fitted Emission Lines Spectra}
\label{sii_fitted_spectra}
We display here a few examples of background-subtracted and fitted \sii\text{} emission line doublet spectra for various regions of the filamentary nebula surrounding NGC 1275 considered in this paper.

\begin{figure*}[h!]
  \centering
  \includegraphics[width=0.80\linewidth]{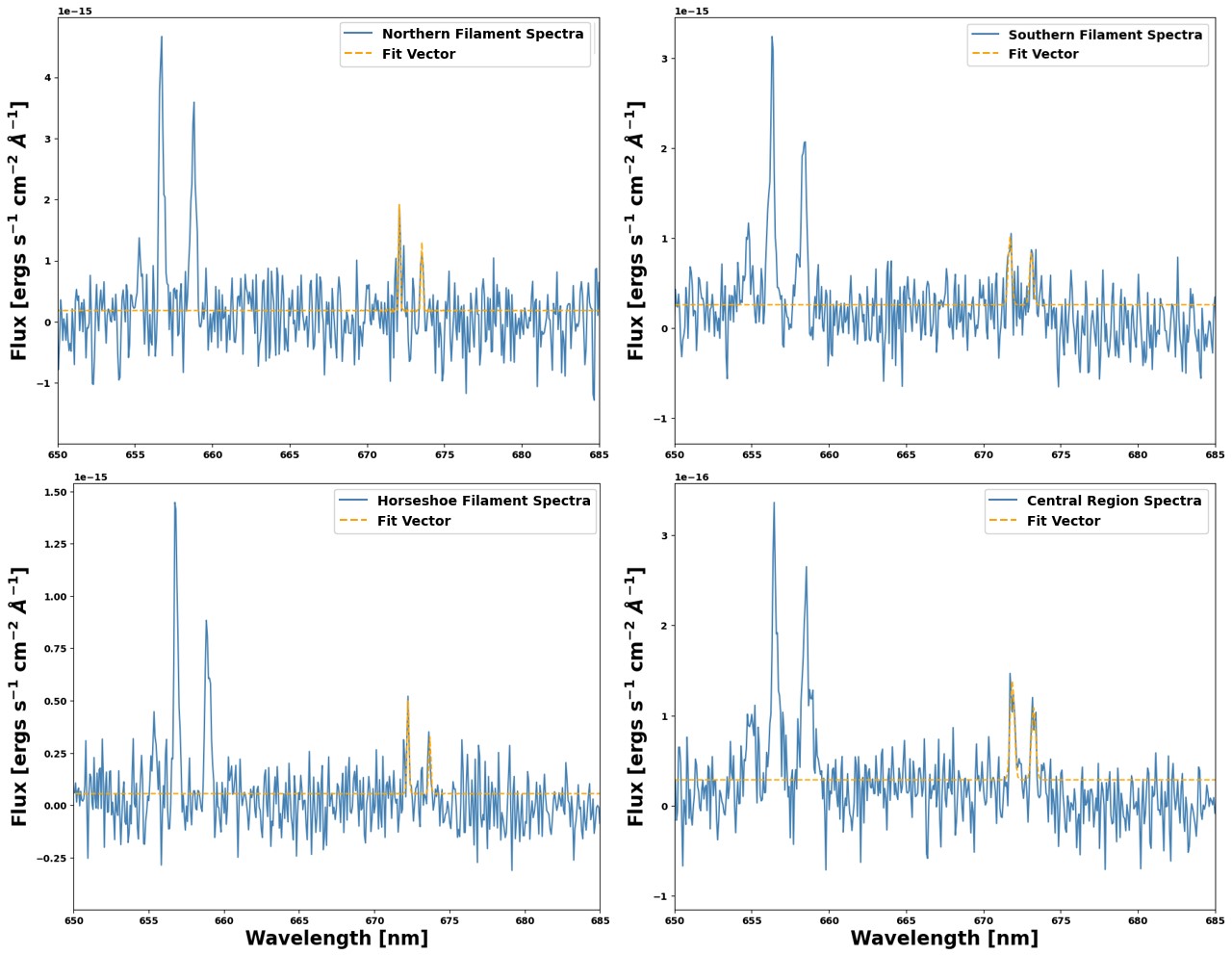}
\caption{Examples of background-subtracted and fitted \sii\text{} emission line doublet for various regions across the filamentary nebula surrounding NGC 1275, namely in the northern filaments (top left), southern filaments (top right), Horseshoe filament (bottom left) and a weighted Voronoi bin of the Central structure (bottom right).}
\label{anx:sii_fitted_spectra}
\end{figure*}



\bibliography{sample631}{}
\bibliographystyle{aasjournal}


\end{document}